\documentclass{rrparticle}
\usepackage{graphicx}
\usepackage[dvips]{epsfig}

\title{Analytical results on quantum correlations of few bosons in a double-well trap} 
\author[1]{Mario Galante}
\author[1,2,a]{Giovanni Mazzarella}
\author[1,2,3]{Luca Salasnich}
\affil[1]{Dipartimento di Fisica e Astronomia ``Galileo Galilei'',Via Marzolo 8, 35131-Padova, Italy}
\affil[2]{CNISM, Consorzio Interuniversitario per le Scienze Fisiche della Materia, Via Marzolo 8, 35131-Padova, Italy} 
\affil[3]{Istituto Nazionale di Ottica (INO) del Consiglio Nazionale delle Ricerche (CNR)\\Sezione di Sesto Fiorentino, Via Nello Carrara 1, 50019 Sesto Fiorentino, Italy\\
Email:$^a$ {\em giovanni.mazzarella@unipd.it}}


\begin{document}
\maketitle
\begin{abstract}
We consider a finite number $N$ of interacting bosonic atoms at zero temperature confined in a one-dimensional double-well trap and study this system by using the two-site Bose-Hubbard (BH) Hamiltonian. For systems with $N=2$ and $N=3$, and $N=4$ bosons we analytically solve the eigenproblem associated to this Hamiltonian and find its lowest energetic state. We investigate the structure of the ground state by varying the strength of the boson-boson interaction from the strongly attractive regime to the deep repulsive one. We characterize the ground state of the two-site BH Hamiltonian by calculating the Fisher information $F$, the coherence visibility $\alpha$, and the entanglement entropy $S$. For these quantities we provide analytical formulas that we use to study $F$, $\alpha$, and $S$  as functions of the interaction between the particles.  We discuss the difference existing, in the deep repulsive regime, between the case with an even number of bosons and that with an odd number of particles, both in the structure of the lowest energetic state and in the behavior of the three above ground-state characterizing parameters.
\end{abstract}

\section{Introduction}

Ultracold and interacting dilute alkali-metal vapors trapped by one-dimensional double-well potentials \cite{oliver}
provides the possibility of studying the formation of macroscopic coherent states \cite{smerzi,stringari,anglin,mahmud,anna}
and macroscopic Schr\"odinger-cat states \cite{cirac,dalvit,huang,carr,brand,main}.
The two-site Bose-Hubbard (BH) Hamiltonian \cite{twomode:milburn} efficiently describes the microscopic
dynamics of such systems. When the boson-boson interaction is repulsive and the number of bosons is even, the crossover from a delocalized atomic coherent
state to a (fully incoherent) localized Fock state (the so called twin Fock state with the particles equally shared between the two wells) takes place by increasing the interatomic coupling strength
\cite{stringari,anglin,mahmud,anna,main}. For attractively interacting bosons, the two-spatial mode BH model predicts the formation of a macroscopic Schr\"odinger-cat state
\cite{cirac,dalvit,huang,carr,brand} when the interatomic attraction becomes sufficiently large. Finally, when the attraction between
the bosons is sufficiently strong the collapse should take place \cite{sb,io-e-boris}.

Motivated by the concrete possibility to isolate single atomic ions \cite{heidelberg,wineland1,bergquist,cheinet} and manipulate quantum gases at single-atom level \cite{cheinet, diedrich, monroe,haroche-raimond,nobel} (note that D. J. Wineland was awarded in 2012 with the physics Nobel prize for his activity in this sector), we focus on the behavior of few trapped bosonic atoms at zero temperature.

The aim of the present work, then, is to study the ground state of a system consisting of a low number $N$ of bosons confined in a symmetric double-well trap and characterize it from the quantum correlations point of view. To do this we use the two-site Bose-Hubbard model. We diagonalize the underlying Hamiltonian by analytically finding the eigenvector and the eigenvalue of its lowest energetic state for $N=2$ - this case has already been discussed in \cite{main} - and $N=3,4$ bosons. Hence, we provide analytical formulas for the parameters that describe the correlation properties of the ground state of the system. These parameters are: the Fisher information $F$ \cite{braunstein,pezze} which is related to the fluctuation of the number of bosons in a given well and achieves its maximum in correspondence to the Schr\"odinger-cat state; the coherence visibility $\alpha$ \cite{stringari,anglin,anna} which measures the coherence related to the single-particle tunneling across the central barrier and attains its maximum value in correspondence to the atomic coherent state; the entanglement entropy $S$ \cite{bwae} which quantifies the amount of the genuine quantum correlations of the ground state from the bi-partition perspective. In particular, we calculate $F$ and $\alpha$ following two paths: on one hand by taking the average, with respect to the ground state, of the left-right population imbalance variance and the left-well hopping operator, respectively, and on the other hand by applying the Hellmann-Feynman theorem \cite{cohen}. For both the calculations (that, as we shall comment, provide the same results) we use the analytically determined ground-state eigenvectors and eigenvalues.

We study the ground state and the parameters $F$, $\alpha$, $S$ by widely exploring the atom-atom interaction range, from strong attractions to strong repulsions. In this latter regime, we comment about the of $N$ even-$N$ odd  difference: when $N$ is even (the ratio of the number of bosons to the number of wells is a positive integer) the ground state is a separable Fock state  with $N/2$ particles in the left well and $N/2$ particles in the right well (this is, as commented at the beginning, the twin Fock state), while when $N$ is odd (the total number of bosons is not commensurate with the number of wells) the ground state is given by a symmetric combination of two separable Fock states. When the boson-boson repulsion becomes sufficiently large, the quantities $F$, $\alpha$, $S$, tend to zero for an even number of particles; they remain, instead, finite when $N$ is odd.

\section{The system}

We analyze a finite number $N$ of identical interacting bosonic atoms at zero temperature confined by a trapping potential $V_{trap}(\bf{r})$. We suppose that this potential is given by the superposition of an isotropic harmonic confinement in the radial plane ($x-y$) and a double-well potential $V_{DW}(z)$ in the axial ($z$) direction, i.e.
\begin{equation}
\label{trapping}
V_{trap}({\bf r}) = \frac{m\omega_{\bot}^2}{2}(x^2+y^2)+ V_{DW}(z)
\;,\end{equation}
where $m$ is the mass of the bosons and $\omega_{\bot}$ the trapping frequency in the radial plane. We assume that the double-well is symmetric in the $z$ direction and that the system is quasi one-dimensional due to a strong transverse radial harmonic confinement.

%\begin{figure}[!h]
%\begin{center}
% \includegraphics[width=.50\columnwidth]{DW.pdf}
%\caption{Shape of the double well potential $V_{DW}(z)$.}
%\end{center}
%\end{figure}

In the second quantization language, the Hamiltonian that controls the microscopic dynamics of the system is
\begin{eqnarray}
\label{system:ham0}
\hat{H} &=& \int d^{3}{\bf r}\hat{\Psi}^\dagger({\bf r})(-\frac{\hbar^2}{2m}\nabla^2+V_{trap}({\bf r}))\hat{\Psi}({\bf r}) \nonumber\\
&+&\frac{1}{2}\int d^{3} {\bf r}d^{3} {\bf r'}\hat{\Psi}^\dagger({\bf r})\hat{\Psi}^\dagger({\bf r'}) V({\bf r}-{\bf r'})
\hat{\Psi}({\bf r'})\hat{\Psi}({\bf r})
\;.\end{eqnarray}
The field operator $\hat{\Psi}({\bf r})$ ($\hat{\Psi}^\dagger({\bf r})$) destroys (creates) a boson in the position ${\bf r}$. $\hat{\Psi}({\bf r})$ and $\hat{\Psi}^\dagger({\bf r})$ satisfy the usual bosonic commutation rules: $[\hat{\Psi}({\bf r}),\hat{\Psi}^\dagger({\bf r'})]=\delta^{(3)}({\bf r}-{\bf r'})$, and $[\hat{\Psi}({\bf r}),\hat{\Psi}({\bf r'})]=0=[\hat{\Psi}({\bf r})^\dagger,\hat{\Psi}^\dagger({\bf r'})]$.
We assume that the bosons interact between each other via short-range interactions, so that the atom-atom interaction potential $V({\bf r}-{\bf r'})$  can be described (in the dilute regime and for ultra-low temperatures) by a contact potential given by
\begin{equation}
\label{contact}
V({\bf r}-{\bf r'})=g\delta^{(3)}({\bf r}-{\bf r'})
\;,\end{equation}
where the coupling constant $g$ is equal to $\displaystyle{\frac{4\pi\hbar a_s}{m}}$ with $a_s$ the s-wave scattering length.
Therefore the Hamiltonian (\ref{system:ham0}) becomes
\begin{eqnarray}
\label{system:ham1}
\hat{H} &=& \int d^{3}{\bf r}\hat{\Psi}^\dagger({\bf r})(-\frac{\hbar^2}{2m}\nabla^2+V_{trap}({\bf r}))\hat{\Psi}({\bf r}) \nonumber\\
&+&\frac{g}{2}\int d^{3} {\bf r}d^{3} {\bf r'}\hat{\Psi}^\dagger({\bf r})\hat{\Psi}^\dagger({\bf r'})
\hat{\Psi}({\bf r'})\hat{\Psi}({\bf r})
\;.\end{eqnarray}
Under the hypothesis that only the lowest energetic doublet of the potential $V_{DW}(z)$ is populated, we expand the field operator $\hat{\Psi}({\bf r})$ according the two-spatial mode decomposition:
\begin{equation}
\label{expansion}
\hat{\Psi}({\bf r}) = \Phi_L({\bf r})\hat{a}_L + \Phi_R({\bf r})\hat{a}_R
\;,\end{equation}
where $\hat{a}_k$ ($\hat{a}^\dagger_k$) - $k=L,R$, with $L (R)$ denoting the left (right) well - destroys (creates) a boson in the $k$th well. The single-particle operators $\hat{a}_k$ and $\hat{a}^\dagger_k$  satisfy the
bosonic commutation rules:
\begin{eqnarray}
\label{spcommutation}
&&[\hat{a}_k,\hat{a}^\dagger_j]=\delta_{k,j} \nonumber\\
&&[\hat{a}_k,\hat{a}_j]=0=[\hat{a}^\dagger_k,\hat{a}^\dagger_j]
\;.\end{eqnarray}
Due to the form  of the trapping potential given by Eq. (\ref{trapping}), the single-particle wave function  $\Phi_k({\bf r})$ ($k=L,R$) can be written according to the factorization
\begin{equation}
\label{system:dec}
\Phi_k({\bf r}) = w(x)w(y)\phi_k(z)
\;,\end{equation}
where $w(x)$ and $w(y)$ are the ground-state wave functions of the harmonic oscillator potentials $m\omega_{\bot}^{2} x^2/2$ and $m\omega_{\bot}^{2} y^{2}/2$, respectively.
The single-particle wave functions $\phi_L(z)$ and $\phi_R(z)$ are tightly localized in the left and right well, respectively, and satisfy the orthonormalization conditions $(k,l = L,R)$
\begin{eqnarray}
\label{system:rel1}
&&\int_{-\infty}^{+\infty}dz|\phi_k(z)|^2 = 1 \nonumber\\
&&\int_{-\infty}^{+\infty}dz \phi^{*}_{k}(z)\phi_l(z) =\delta_{k,l}
\;,\end{eqnarray}
(with $\phi^{*}_{k}(z)$ the complex conjugate of $\phi_{k}(z)$) so that
\begin{eqnarray}
\label{system:rel2}
&&\int d^3 {\bf r}|\Phi_k({\bf r})|^2 =1\nonumber\\
&&\int d^3 {\bf r} \Phi^{*}_{k}({\bf r})\Phi_l({\bf r})=\delta_{k,l}
\;.\end{eqnarray}
We use the expansion (\ref{expansion}) and its Hermitian conjugate at the right-hand side of Eq. (\ref{system:ham1}); by exploiting the orthonormalization conditions
(\ref{system:rel2}) and the fact that $V_{DW}(z)$ is symmetric, the well known two-site Bose-Hubbard Hamiltonian \cite{main,twomode:milburn} is achieved
\begin{equation}
\label{ham:bh}
\hat{H} = -J(\hat{a}_L^\dagger\hat{a}_R + \hat{a}_R^\dagger\hat{a}_L) +\frac{U}{2}\big(\hat{n}_L(\hat{n}_L-1)+\hat{n}_R(\hat{n}_R-1)\big)
\;.\end{equation}
Here $\hat{n}_k = \hat{a}^\dagger_k\hat{a}_k$ is the operator counting the number of bosons in the $k$th well. Note that the Hamiltonian (\ref{ham:bh}) commutes with the total number operator $\hat{N}=\hat{n}_L+\hat{n}_R$. The amplitude $U$ measures the strength of the boson-boson interaction in the same well (on-site or intra-well interaction)
\begin{equation}
\label{u}
U =\frac{g}{2\pi a_\perp^2}\int_{-\infty}^{+\infty}dz |\phi_k(z)|^4
\end{equation}
with $a_\bot =\displaystyle{\sqrt{\frac{\hbar}{m\omega_{\bot}}}}$. The sign of $U$ is controlled by that of $a_s$ which can be experimentally tuned via the Feshbach resonance technique, so that when $a_s$ is positive (negative) the bosons are repulsively (attractively) interacting. $J$ is the tunnel matrix element between the two wells:
\begin{equation}
\label{j}
J =-\int_{-\infty}^{+\infty} dz\phi^{*}_{L}(z)\,\big(-\frac{\hbar^2}{2m}\frac{d^2}{dz^2}+V_{DW}(z)\big)\,\phi_R(z)
\;.\end{equation}

To capture the main properties of the system, we focus on the eigenproblem
\begin{equation}
\label{eigenproblem}
\hat {H} |E_j\rangle = E_j |E_j \rangle
\end{equation}
for a fixed number $N$ of bosons. In this case the Hamiltonian $\hat {H}$ can be represented by
a $(N+1)\times(N+1)$ matrix in the Fock basis $|i,N-i\rangle=|i\rangle_L \otimes |N-i\rangle_R$ (with $\otimes$ denoting the tensor product) with $i=0,...,N$. For each eigenvalue $E_j$, with
$j=0,1,...,N$, the corresponding eigenstate $|E_j\rangle$ will be of the form
\begin{equation}
\label{eigenstate}
|E_j\rangle=\sum_{i=0}^{N}\,c_{i}^{(j)} \, |i,N-i\rangle \;,
\end{equation}
where $|c_i^{(j)}|^2$ is the probability to have $i$ ($N-i$) bosons in the left (right) well when the system is in the $j$th eigenstate of the two-site BH Hamiltonian.
Note that since the left-right symmetry of the Hamiltonian (\ref{ham:bh}), for any eigenstate one has that
\begin{equation}
\label{lrs}
\langle \hat{n}_L \rangle=\langle \hat{n}_R \rangle
\;,\end{equation}
where the average $\langle...\rangle$ is taken with respect to the given eigenstate.
We are analyzing the system at zero temperature. Then, the only two-site BH Hamiltonian eigenstate to be occupied is the lowest one, so that in the following we shall denote the corresponding eigenvector and eigenvalue simply by $|E\rangle$ and $E$, respectively. The expansion coefficients with respect to the basis $|i,N-i\rangle$ shall be, then, denoted by $c_i$. As discussed in \cite{main}, the ground state of the Hamiltonian (\ref{ham:bh}) features different behaviors depending on the interplay between the on-site interaction $U$ and the hopping amplitude $J$. Following the same path followed in \cite{main}, we study the ground state in terms of the dimensionless parameter $\xi =U/J$. Let us start with some limit cases.
\begin{itemize}
\item $\xi=0$. The ground state is the atomic coherent state \cite{arecchi}
\begin{equation}
\label{ACS}
|ACS\rangle = \frac{1}{\sqrt{N!}}(\frac{1}{\sqrt{2}}\big(\hat{a}_L^\dagger+\hat{a}_R^\dagger)\big)^N|0,0\rangle
\;,\end{equation}
(the energy associated to this state is $-NJ$) where $|0,0\rangle=|0\rangle_L\otimes|0\rangle_R$ is the tensor product between the vacuum of the operator $\hat{a}_L$ and the vacuum of $\hat{a}_R$, i.e. we have no particles in the left well and no particles in the right well.
\item $U>0$ : $\xi\rightarrow+\infty$. In the case of a strong repulsive interaction and with an even number $N$ of bosons, as well known, the ground state tends to the twin Fock state
\begin{equation}
\label{fock}
|FOCK\rangle = |\frac{N}{2},\frac{N}{2}\rangle
\;.\end{equation}
%To prove this, let us suppose that $J=0$ and consider the eigenvalues $n_L$ and $n_R$ of the  operators $\hat{n}_L$ and $\hat{n}_R$, which we denote, respectively,
%with $n_L=x$ and $n_R=N-x$ sicne the number of particles in the system is fixed. Then, the energy $E$ will be a function of $x$ that we can write (see Eq. (\ref{ham:bh})) as
%\begin{equation}
%E_0 = \frac{U}{2}(x(x-1)+(N-x)(N-x-1)) = \frac{U}{2}(2x^2-2Nx+N^2-N),
%\end{equation}
%with $U>0$. Thus, deriving $E$ with respect to $x$, shows that $E$ achieved its minimum when $x=N/2$.\\

If $N$, instead, is odd, when  $\xi\rightarrow+\infty$ the ground state tends to
%\begin{widetext}
\begin{equation}
\label{pseudo:fock}
|pseudoFOCK\rangle = \frac{1}{\sqrt{2}}\Big(\Big|\frac{N-1}{2},\frac{N+1}{2}\Big\rangle + \Big|\frac{N+1}{2},\frac{N-1}{2}\Big\rangle\Big)
\;.\end{equation}
%\end{widetext}
To understand this, let us consider the extreme case of complete absence of hopping, that is $J=0$. In this case the eigenvalues of the Hamiltonian (\ref{ham:bh}) are given by those of the intra-well term: $E=\displaystyle{\frac{U}{2}(2i^2-2Ni+N^2-N)}$. We are here considering the state with $i$ $(N-i)$ bosons in the left (right) well. Requiring that $\partial E/\partial i=0$ provides $i=N/2$. Since $N$ is odd and $i$ must be an integer, the values of $i$ which minimize $E$ are those integer closest to $N/2$, i.e. $i=(N-1)/2$ and $i=(N+1)/2$ that correspond to the two separable Fock states
\begin{eqnarray}
&&|\varphi\rangle_1 = |\frac{N-1}{2},\frac{N+1}{2}\rangle \nonumber\\
&&|\varphi\rangle_2 = |\frac{N+1}{2},\frac{N-1}{2}\rangle
\;.\end{eqnarray}
These states, although having the (same) minimum energy, do not satisfy the condition (\ref{lrs}). Nevertheless, it is easy to prove that the state
\begin{equation}
|\varphi\rangle_3 = \frac{1}{\sqrt{2}}(|\varphi\rangle_1+|\varphi\rangle_2)
\end{equation}
has the same energy of $|\varphi\rangle_l$ ($l=1,2$) and satisfies the condition (\ref{lrs}).
\item $U<0$ : $\xi\rightarrow-\infty$. In the case of a strong attractive interaction, the ground state tends to the macroscopic superposition state
\begin{equation}
\label{cat}
|CAT\rangle=\frac{1}{\sqrt{2}}(|N,0\rangle+|0,N\rangle)
\;.\end{equation}
\end{itemize}
This state, frequently called NOON state, is the boson-version of the Schr\"odinger cat state \cite{cirac,dalvit,huang,carr,brand}.

At this point it is worth to observe that apart the issue of the possible collapse (related to attractive interactions), the realization of the cat state is not trivial due to the very tiny separation (in the presence of finite couplings) between the two lowest levels that makes the cat state very fragile, see, for example, \cite{huang}.

\section{Analysis parameters}
In this section we introduce the parameters that we use to characterize the correlations of the ground state of the two-site BH Hamiltonian (\ref{ham:bh}). These parameters are the Fisher information, the coherence visibility, and the entanglement entropy.

In the meanwhile, it is useful to remind the well-known properties:
\begin{equation}
\label{ort}
\langle j,N-j|i,N-i\rangle=\delta_{i,j}
\;,\end{equation}
and
\begin{eqnarray}
\label{spaction}
&&\hat{a}^\dagger_L |i,N-i\rangle = \sqrt{i+1}|i+1,N-i\rangle\nonumber\\
&&\hat{a}_L |i,N-i\rangle = \sqrt{i}|i-1,N-i\rangle\nonumber\\
&&\hat{a}^\dagger_R |i,N-i\rangle = \sqrt{N-i+1}|i,N-i+1\rangle\nonumber\\
&&\hat{a}_R |i,N-i\rangle = \sqrt{N-i}|i,N-i-1\rangle\nonumber\\
&&\hat{n}_L|i,N-i\rangle = i|i,N-i\rangle. \nonumber\\
&&\hat{n}_R|i,N-i\rangle = (N-i)|i,N-i\rangle
\;.\end{eqnarray}
\begin{itemize}
\item {\it Fisher Information.}\\
The quantum Fisher information $F_{QFI}$ is the quantity \cite{braunstein,pezze}
\begin{equation}
\label{qfi}
F_{QFI} =(\Delta\hat{n}_{L,R})^2 = \langle(\hat{n}_L-\hat{n}_R)^2\rangle - (\langle\hat{n}_L-\hat{n}_R\rangle)^2
\;,\end{equation}
where the expectation values are taken with respect to the ground state $|E\rangle$.
By using  the orthonormality condition (\ref{ort}) and rules (\ref{spaction}) in Eq. (\ref{qfi}), we can express $F_{QFI}$ in terms of the expansion coefficients $c_i$ as follows:
\begin{equation}
F_{QFI} = \sum_{i=0}^N (2i-N)^2|c_i|^2
\;.\end{equation}
It is convenient to normalize $F_{QFI}$ at its maximum value $N^2$ by defining the Fisher information $F$ as
\begin{equation}
\label{fi}
F=\frac{F_{QFI}}{N^2}
\;,\end{equation}
so that we have a quantity varying in the range $[0,1]$. In terms of the coefficients $c_i$, $F$ is
\begin{equation}
\label{fi2}
F=\frac{1}{N^2}\sum_{i=0}^N (2i-N)^2|c_i|^2
\;.\end{equation}
This $F$ will be equal to $1$ for the NOON state (\ref{cat}).

\item {\it Coherence visibility.}\\
In ultracold atom physics, it is customary to investigate the coherence properties in terms of the momentum distribution $n(p)$ which is the Fourier transform of the one-body
density matrix $\rho_1(x,x')$ \cite{stringari,anglin,anna}:
\begin{equation}
\label{np}
n(p)=\int dx dx'\exp\big(-ip(x-x')\big)\,\rho_1(x,x') \;,
\end{equation}
where
\begin{equation}
\rho_1(x,x')=\langle \hat{\Psi}(x)^{\dagger}\hat{\Psi}(x')\rangle \;
\end{equation}
with the operators $\hat{\Psi}(x)$ and $\hat{\Psi}^{\dagger}(x)$ -
satisfying the standard bosonic commutation rules - annihilating and creating,
respectively, a boson at the point $x$, and the average $\langle...\rangle$ being the ground-state average.
Following Refs. \cite{stringari,anglin,anna}, it is possible to show that the momentum distribution $n(p)$ can be written as
\begin{equation}
\label{npexp}
n(p)= n_0(p) \bigg(1 +\alpha \cos\big(pd \big) \bigg) \;.
\end{equation}
Here $n_0(p)$ is the momentum distribution in the fully incoherent regime
($n_0(p)$ depends on the shape of the double-well potential $V_{DW}(z)$), and
$d$ is the distance between the two minima of $V_{DW}(z)$. $\alpha$ is
a real quantity which measures the visibility of the interference fringes. This visibility is given by
\begin{equation}
\label{visibility}
\alpha=\frac{2\,|\langle \hat{a}^{\dagger}_L\hat{a}_R\rangle|}{N} \;,
\end{equation}
where the expectation value is taken with respect to the ground state. The quantity $\alpha$ characterizes the degree of coherence, between the two wells, related to the left-right (and back) tunneling.

We can express the  coherence visibility (\ref{visibility}) in terms of the coefficients $c_i$ by using in Eq. (\ref{visibility}) the rules (\ref{spaction}) and the orthonormalization condition (\ref{ort}), so that one has
\begin{equation}
\label{visibility2}
\alpha = \frac{2}{N}|\sum_{i=0}^N c_{i} c^{*}_{i+1} \sqrt{(i+1)(N-i)}|
\;,\end{equation}
where $c^{*}_{i+i}$ is the complex conjugate of $c_{i+1}$. $\alpha$ is maximum, that is $1$, for the atomic coherent state (\ref{ACS}).

\item {\it Entanglement entropy.}\\
Finally, it is interesting to analyze the genuine quantum correlations pertaining to the ground state $|E\rangle$. In particular, we study the quantum entanglement
of $|E\rangle$ from the perspective of the bi-partition. In this framework, the two partitions are given by the left well and right one.  When the system is in $|E\rangle$, the
density matrix $\hat{\rho}$ is
\begin{equation}
\label{dm}
\hat{\rho} =|E\rangle\langle E|
\;.\end{equation}
An excellent measure of the entanglement between the two wells
is provided by the entanglement entropy $S$ \cite{bwae}. This quantity is the von Neumann entropy of the reduced density matrix $\hat{\rho}_{L(R)}$
defined by
\begin{equation}
\hat{\rho}_{L(R)} =Tr_{R(L)} \hat{\rho}
\;,\end{equation}
that is the matrix obtained by partial tracing the total density matrix (\ref{dm}) over the degrees of freedom of the right (left) well (note that $\hat{\rho}_{L}=\hat{\rho}_{R}$). By using the definition of trace of a matrix and the orthonormalization condition (\ref{ort}), it is possible to show that the entanglement entropy
\begin{equation}
\label{ee0}
S=-Tr \hat{\rho}_{L(R)} \log_{2}\hat{\rho}_{L(R)}
\end{equation}
is given by
\begin{equation}
\label{ee}
S=-\sum_{i=0}^{N}|c_{i}|^2\log_{2}|c_{i}|^2
\;.\end{equation}
For a given number of bosons $N$, the theoretical maximum value of $S$ is $\log_2(N+1)$ that would correspond to the situation in which the quantities $|c_{i}|^2$ are all equal: $|c_i|^2=1/(N+1)$ whatever $i$.
\end{itemize}
\begin{center}
\begin{tabular}{|c|c|c|c|c|}
\hline
~~~$State (N=2)$~~~ & ~~~$F$~~~ &~~~$\alpha$~~~ & ~~~$S$~~~\\
\hline
$|ACS\rangle$
%$\displaystyle{\bigg(\frac{1}{2^{N}}\frac{N!}
%{i!(N-i)!}\bigg)^{\frac{1}{2}}} $
& $1/2$
& $1$ & $3/2$  \\
$|FOCK\rangle$
%$1\, \mbox{for}\, \displaystyle{i=\frac{N}{2}}; 0 \,\mbox{otherwise} $
%$\delta_{i,N/2}$
&$0$ & $0$ & $0$ \\
$|CAT\rangle$
%$\displaystyle{\frac{1}{\sqrt{2}}}
%\,\mbox{for}\, i=0,N; 0 \,\mbox{otherwise}$
%\left [ \delta_{i,0}+\delta_{i,N}\right ]$
&$1$ & $0$ & $1$\\
\hline
\end{tabular}
\end{center}
Table 1. {\small The Fisher information $F$, the coherence visibility $\alpha$, and the entanglement entropy $S$, for the atomic coherent state (\ref{ACS}), the twin Fock state (\ref{fock}), and the NOON state (\ref{cat}) with $N=2$ bosons.}\\

\begin{center}
\begin{tabular}{|c|c|c|c|c|}
\hline
~~~$State (N=3)$~~~ & ~~~$F$~~~ &~~~$\alpha$~~~ & ~~~$S$~~~\\
\hline
$|ACS\rangle$
%$\displaystyle{\bigg(\frac{1}{2^{N}}\frac{N!}
%{i!(N-i)!}\bigg)^{\frac{1}{2}}} $
& $1/3$
& $1$ & $1.81128$  \\
$|pseudoFOCK\rangle$
%$1\, \mbox{for}\, \displaystyle{i=\frac{N}{2}}; 0 \,\mbox{otherwise} $
%$\delta_{i,N/2}$
&$1/9$ & $2/3$ & $1$ \\
$|CAT\rangle$
%$\displaystyle{\frac{1}{\sqrt{2}}}
%\,\mbox{for}\, i=0,N; 0 \,\mbox{otherwise}$
%\left [ \delta_{i,0}+\delta_{i,N}\right ]$
&$1$ & $0$ & $1$\\
\hline
\end{tabular}
\end{center}
%\end{widetext}
Table 2. {\small The Fisher information $F$, the coherence visibility $\alpha$, and the entanglement entropy $S$, for the atomic coherent state (\ref{ACS}), the state (\ref{pseudo:fock}), and the NOON state (\ref{cat}) with $N=3$ bosons.}\\

\begin{center}
\begin{tabular}{|c|c|c|c|c|}
\hline
~~~$State (N=4)$~~~ & ~~~$F$~~~ &~~~$\alpha$~~~ & ~~~$S$~~~\\
\hline
$|ACS\rangle$
%$\displaystyle{\bigg(\frac{1}{2^{N}}\frac{N!}
%{i!(N-i)!}\bigg)^{\frac{1}{2}}} $
& $1/4$
& $1$ & $2.03064$  \\
$|FOCK\rangle$
%$1\, \mbox{for}\, \displaystyle{i=\frac{N}{2}}; 0 \,\mbox{otherwise} $
%$\delta_{i,N/2}$
&$0$ & $0$ & $0$ \\
$|CAT\rangle$
%$\displaystyle{\frac{1}{\sqrt{2}}}
%\,\mbox{for}\, i=0,N; 0 \,\mbox{otherwise}$
%\left [ \delta_{i,0}+\delta_{i,N}\right ]$
&$1$ & $0$ & $1$\\
\hline
\end{tabular}
\end{center}
Table 3. {\small The Fisher information $F$, the coherence visibility $\alpha$, and the entanglement entropy $S$, for the atomic coherent state (\ref{ACS}), the twin Fock state (\ref{fock}), and the NOON state (\ref{cat}) with $N=4$ bosons.}\\

\section{Analysis}
In this section, we determine the ground state of the two-site Bose-Hubbard Hamiltonian when $N=1$, $N=2$, $N=3$, and $N=4$. We calculate the Fisher information (\ref{fi}), the coherence visibility (\ref{visibility}), and the entanglement entropy (\ref{ee}) for a system with $N=1$, $N=2$, $N=3$, and $N=4$ bosons. We analyze the structure of the ground state and $F$, $\alpha$, $S$  in terms of the scaled on-site interaction $\xi=U/J$.

As first, we represent the Hamiltonian $\hat{H}$ with respect to the Fock basis $|i,N-i\rangle$. We start from the right-hand side of Eq. (\ref{ham:bh}) and use the rules (\ref{spaction}) and the orthonormalization condition (\ref{ort}). Note that, here, we measure the energies in units of $J$. We shall denote by the symbols $\hat{\tilde H}$ and $\tilde E$ the dimensionless energetic quantities, i.e. $\hat{\tilde H}=\hat{H}/J$ and $\tilde E=E/J$.

When $N=1$, the Hamiltonian (\ref{ham:bh}) consists of the only hopping term. In this case, the two-site Bose-Hubbard Hamiltonian, given by Eq. (\ref{ham:bh}), in the Fock basis $|i,N-i\rangle$ is
\[\hat{\tilde H}= \left(\begin{array}{cc}
  0  & -1 \\
  -1 & 0 \end{array}
   \right).\]
The eigenvector $|E\rangle$ associated to the ground state is
\begin{equation}
\label{sgs1}
|E\rangle=\frac{1}{\sqrt{2}}\big(|0,1\rangle+|1,0\rangle\big)
\;,\end{equation}
and the related eigenvalue is
\begin{equation}
\label{e1}
\tilde E=-1
\;.\end{equation}
Then, it is easy to see that the state $|E\rangle$ is the atomic coherent state (\ref{ACS}) with $N=1$ that coincides with the state NOON, Eq. (\ref{cat}), and with the state (\ref{pseudo:fock}) with $N=1$. In this case, by using Eqs. (\ref{fi2}), (\ref{visibility2}), and (\ref{ee}), we immediately see that $F=\alpha=S=1$.

At this point, let us focus on a number of bosons larger than one. We therefore consider the cases $N=2$, $N=3$, and $N=4$. The matrices corresponding to the two-site Bose-Hubbard Hamiltonian (\ref{ham:bh}) with $N=2$, $N=3$, and $N=4$ are, respectively

\[\hat{\tilde H}= \left(\begin{array}{ccc}
  \xi & -\sqrt{2} & 0 \\
 -\sqrt{2} & 0 & -\sqrt{2} \\
  0 & -\sqrt{2} & \xi \end{array} \right),\]

\[\hat{\tilde H}= \left(\begin{array}{cccc}
 3\xi & -\sqrt{3} & 0 & 0 \\
  -\sqrt{3} & \xi & -2 & 0 \\
  0 & -2 & \xi & -\sqrt{3} \\
  0 & 0 & -\sqrt{3} & 3\xi
\end{array} \right),\]

\[\hat{\tilde H}= \left(\begin{array}{ccccc}
 6\xi & -2 & 0 & 0 & 0 \\
 -2 & 3\xi & -\sqrt{6} & 0 & 0 \\
  0 & -\sqrt{6} & 2\xi & -\sqrt{6} & 0 \\
  0 & 0 & -\sqrt{6} & 3\xi & -2 \\
  0 & 0 & 0 & -2 & 6\xi
\end{array} \right),\]

where $\xi=U/J$.

The ground-state energy $\tilde E$ pertaining to the case $N=2$ and the corresponding eigenvector $|E\rangle$ are, respectively
\begin{equation}
\label{egs2}
\tilde E=\frac{1}{2} (\xi - \sqrt{16 + \xi^2})
\;,\end{equation}
\begin{equation}
\label{sgs2}
|E\rangle =  A_2\Big(|0,2\rangle + \frac{\xi + \sqrt{16 + \xi^2}}{2 \sqrt{2}}|1,1\rangle+ |2,0\rangle \Big)
\;,\end{equation}
so that
\begin{eqnarray}
\label{c2}
&&c_0=c_2=A_2\nonumber\\
&&c_1=\frac{A_2(\xi + \sqrt{16 + \xi^2})}{2 \sqrt{2}}
\;.\end{eqnarray}

For $N=3$, we get
\begin{equation}
\label{egs3}
\tilde E = -1 + 2 \xi - \sqrt{4 + 2 \xi + \xi^2}
\;,\end{equation}
\begin{eqnarray}
\label{sgs3}
&&|E\rangle = A_3 \Big( |0,3\rangle + \frac{1 + \xi + \sqrt{4 + 2 \xi + \xi^2}}{\sqrt{3}}|1,2\rangle\nonumber\\
&+& \frac{1 + \xi + \sqrt{4 + 2 \xi + \xi^2}}{\sqrt{3}}|2,1\rangle + |3,0\rangle \Big)
\;,\end{eqnarray}
so that
\begin{eqnarray}
\label{c3}
&&c_0=c_3=A_3\nonumber\\
&&c_1=c_2=\frac{A_3(1 + \xi + \sqrt{4 + 2 \xi + \xi^2})}{\sqrt{3}}
\;.\end{eqnarray}

When $N=4$, for the energy of the ground state we obtain the following result:

\begin{eqnarray}
\label{egs4}
&&\tilde E=\frac{1}{3}\big(11\xi-2\sqrt{k_4}\cos \frac{\theta}{3}\big) \nonumber\\
&&\theta=\theta_1=\arctan \frac{b_4}{a_4} \nonumber\\
&&\theta=\theta_2=\arctan \frac{b_4}{a_4} +\pi
\;,\end{eqnarray}
where $k_4=13 \xi^2+48$. $\theta=\theta_1$ (the second row of Eq. (\ref{egs4})) when $\xi \le-\bar \xi$ and $0<\xi \le \bar \xi$, and $\theta=\theta_2$ (the third row of Eq. (\ref{egs4})) when $-\bar \xi<\xi<0$ and $\xi>\bar \xi$ with $\bar\xi=12\sqrt{2/35}$. Moreover $a_4=288\xi-35\xi^3$, $b_4=6\sqrt{3}\sqrt{9\xi^6+412\xi^4+64 \xi^2+1024}$. Note that when $\xi \rightarrow 0^{+}(^{-})$, the energy in Eq. (\ref{egs4}) gives back $-4$ (in units of $J$) for $\theta=\theta_1$ ($\theta_2$), i.e. the energy of to the atomic coherent state (\ref{ACS}).
The eigenvector pertaining to the energy in Eq. (\ref{egs4}) is
\begin{eqnarray}
\label{sgs4}
&&|E\rangle = A_4 \Big( |0,4\rangle + (3\xi-\frac{\tilde E}{2})|1,3\rangle\nonumber\\
&+& \big(\frac{18\xi^2-9\tilde E\xi+\tilde E^2-4}{2\sqrt{6}}\big)|2,2\rangle \nonumber\\
&+&(3\xi-\frac{\tilde E}{2})|3,1\rangle +|0,4\rangle\Big)
\;,\end{eqnarray}
so that
\begin{eqnarray}
\label{c4}
&&c_0=c_4=A_4\nonumber\\
&&c_1=c_3=A_4(3\xi-\frac{\tilde E}{2})\nonumber\\
&&c_2=\frac{A_4(18\xi^2-9\tilde E\xi+\tilde E^2-4)}{2\sqrt{6}}\;.\nonumber\\
\end{eqnarray}

The factors $A_2$, $A_3$, and $A_4$ are normalization factors given by the following formulas:
\begin{equation}
\label{a2}
A_2 = \frac{2}{\sqrt{16+\xi^2+\xi \sqrt{\xi^2+16}}}
\;,\end{equation}
\begin{equation}
\label{a3}
A_3 = \frac{1}{\sqrt{2 + \frac{2}{3} \big(1 + \xi + \sqrt{4 + \xi (2 + \xi)}\big)^2}}
\;,\end{equation}
%\begin{widetext}
\begin{eqnarray}
\label{a4}
&&A_4=2\sqrt{\frac{6}{d_4}}\nonumber\\
&&d_4=48+12(\tilde E-6\xi)^2+\nonumber\\
&&(18\xi^2-9\tilde E \xi+\tilde E^2-4)^2
\;\end{eqnarray}
%\end{widetext}
with $\tilde E$ given by Eq. (\ref{egs4}). Note that Eqs. (\ref{egs2}) and (\ref{sgs2}) are the same that we found in \cite{main}.
We observe that in the limit $\xi\rightarrow -\infty$, the states (\ref{sgs2}),  (\ref{sgs3}) and (\ref{sgs4})  becomes, as expected,
\begin{equation}
\label{cat2}
|E\rangle = \frac{1}{\sqrt{2}}(|0,2\rangle+|2,0\rangle)
\;,\end{equation}
i.e. the boson-version of the Schr\"odinger cat state (\ref{cat}) with $N=2$,
\begin{equation}
\label{cat3}
|E\rangle = \frac{1}{\sqrt{2}}(|0,3\rangle +|3,0\rangle)
\;,\end{equation}
which is the Schr\"odinger cat state (\ref{cat}) with $N=3$, and similarly
\begin{equation}
\label{cat4}
|E\rangle = \frac{1}{\sqrt{2}}(|0,4\rangle +|4,0\rangle)
\;,\end{equation}

When $\xi\rightarrow 0$, from the state (\ref{sgs2}) ($N=2$) we retrieve
\begin{equation}
\label{acs2}
|E\rangle = \frac{1}{2}(|0,2\rangle+\sqrt{2}|1,1\rangle+|2,0\rangle)
\end{equation}
and from the state (\ref{sgs3}) ($N=3$) one gets
\begin{eqnarray}
\label{acs3}
|E\rangle &=& \frac{1}{2}\Big(\frac{1}{\sqrt{2}}|0,3\rangle + \sqrt{\frac{3}{2}}|1,2\rangle \nonumber\\
 &+&\sqrt{\frac{3}{2}}|2,1\rangle + \frac{1}{\sqrt{2}}|3,0\rangle\Big)
\;.\end{eqnarray}
When $N=4$ and $\xi \rightarrow 0$, the state (\ref{sgs4}) gives
\begin{eqnarray}
\label{acs4}
|E\rangle &=& \frac{1}{4}\Big(|0,4\rangle + 2|1,3\rangle +\sqrt{6}|2,2\rangle \nonumber\\
 &+&2|3,1\rangle +|4,0\rangle\Big)
\;.\end{eqnarray}

These last three states represent the forms assumed by the atomic coherent state $|ACS\rangle$ (\ref{ACS}) when $N=2$, $N=3$, and $N=4$, respectively.

In the limit $\xi\rightarrow +\infty$, the state (\ref{sgs2}) ($N=2$) becomes $|1,1\rangle$ which is the twin Fock state (\ref{fock}) with $N=2$. Instead for $N=3$, in the deep repulsive regime, $\xi\rightarrow +\infty$,  the state (\ref{sgs3}) becomes
\begin{equation}
\label{pf3}
|pseudoFOCK\rangle = \frac{1}{\sqrt{2}}(|1,2\rangle +|2,1\rangle)
\;,\end{equation}
that is the state (\ref{pseudo:fock}) with $N=3$. Finally, when $N=4$ and $\xi\rightarrow +\infty$ , we retrieve the twin Fock state $|2,2\rangle$.

To understand the role of the intra-well interaction-hopping interplay in determining the structure of the ground state of the two-site BH Hamiltonian, we have studied the changes experienced by the probabilities $|c_i|^2$ by varying the scaled on-site interaction $\xi=U/J$ in the presence of $N=2$, $N=3$, and $N=4$ bosons, see Fig. 1. From this figure, we can see that a crossover occurs when $\xi$ ranges from $\xi=-30$ (the two top panels of Fig. 1: the largest probabilities $|c_i|^2$ are located in correspondence to $|0,N\rangle$ and $|N,0\rangle$, this being representative of cat-like states (\ref{cat}) with $N=2,4$ and $N=3$ bosons) to $\xi=30$ (the two bottom panels of Fig. 1: $|c_i|^2$ reaches its largest value in correspondence to  $|1,1\rangle$ when $N=2$ and $|2,2\rangle$ when $N=4$  -  separable twin Fock state (\ref{fock}) - and in correspondence to the states $|1,2\rangle$ and $|2,1\rangle$ when $N=3$, state (\ref{pseudo:fock})) passing for $\xi=0$ (the three panels at the fifth row from the top) describing an almost Gaussian distribution of the probabilities $|c_i|^2$, that is the atomic coherent state (\ref{ACS}). Note that at fifth row of Fig. 1, the plot of $|c_i|^2$ with $N=4$ has been, nominally, labeled by $\xi=0$; actually, this plot has been obtained by performing the limit $\xi \rightarrow 0$ in the equations for $|c_i|^2$ obtained by Eq. (\ref{c4}).

\begin{figure}[htpb]
\centering
{\includegraphics[width=.24\columnwidth]{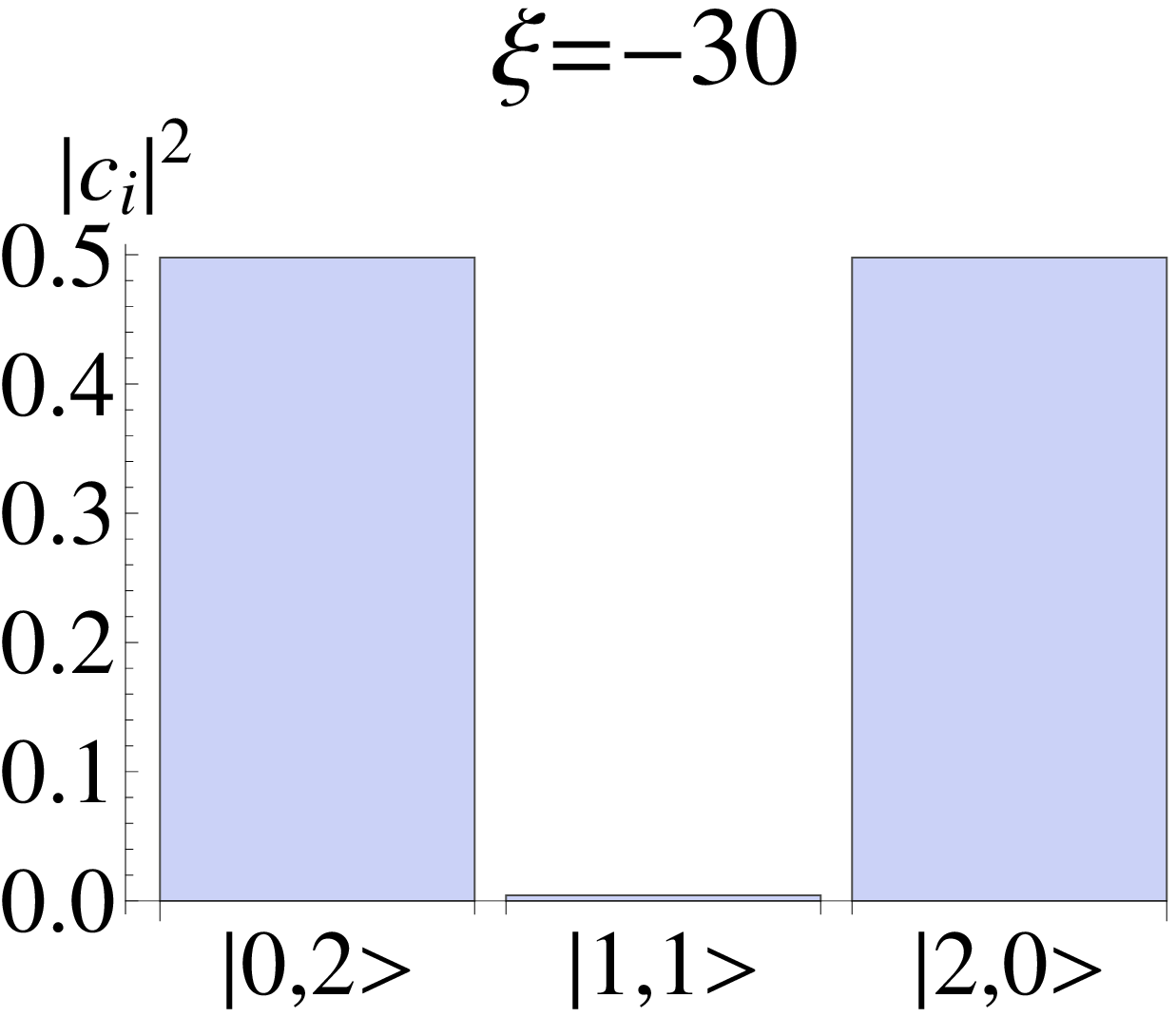}} \quad
{\includegraphics[width=.24\columnwidth]{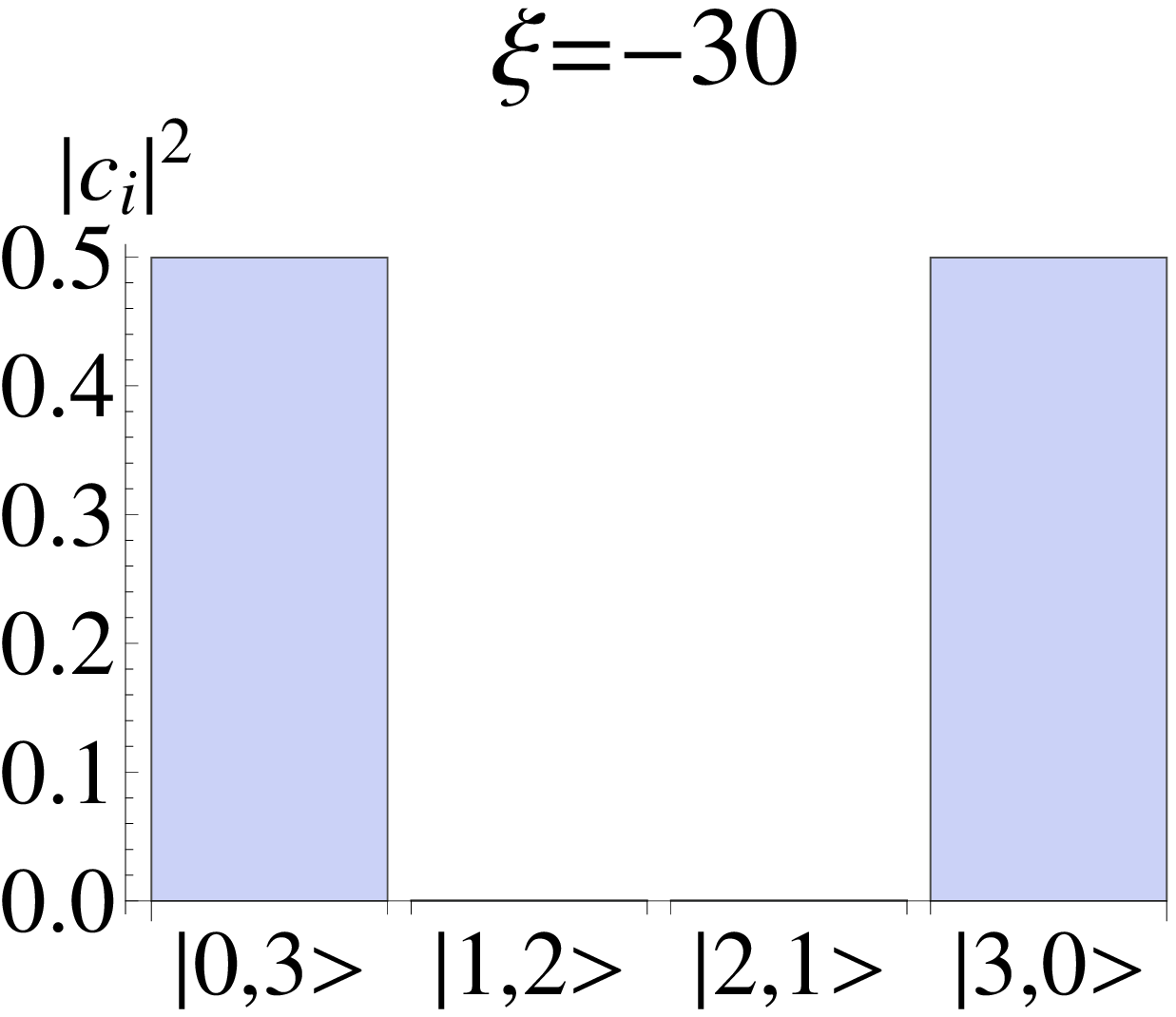}} \quad
{\includegraphics[width=.24\columnwidth]{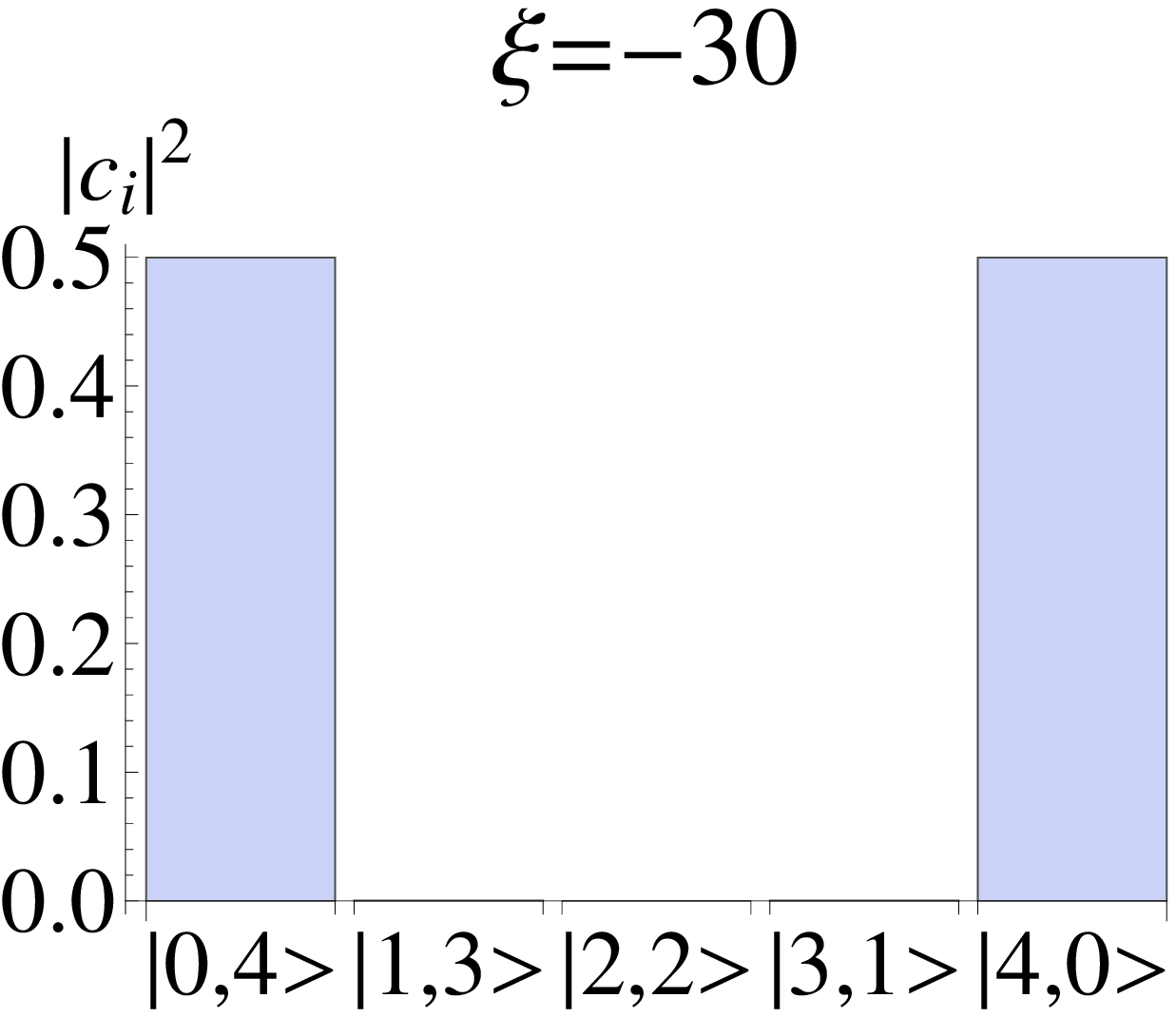}} \\
{\includegraphics[width=.24\columnwidth]{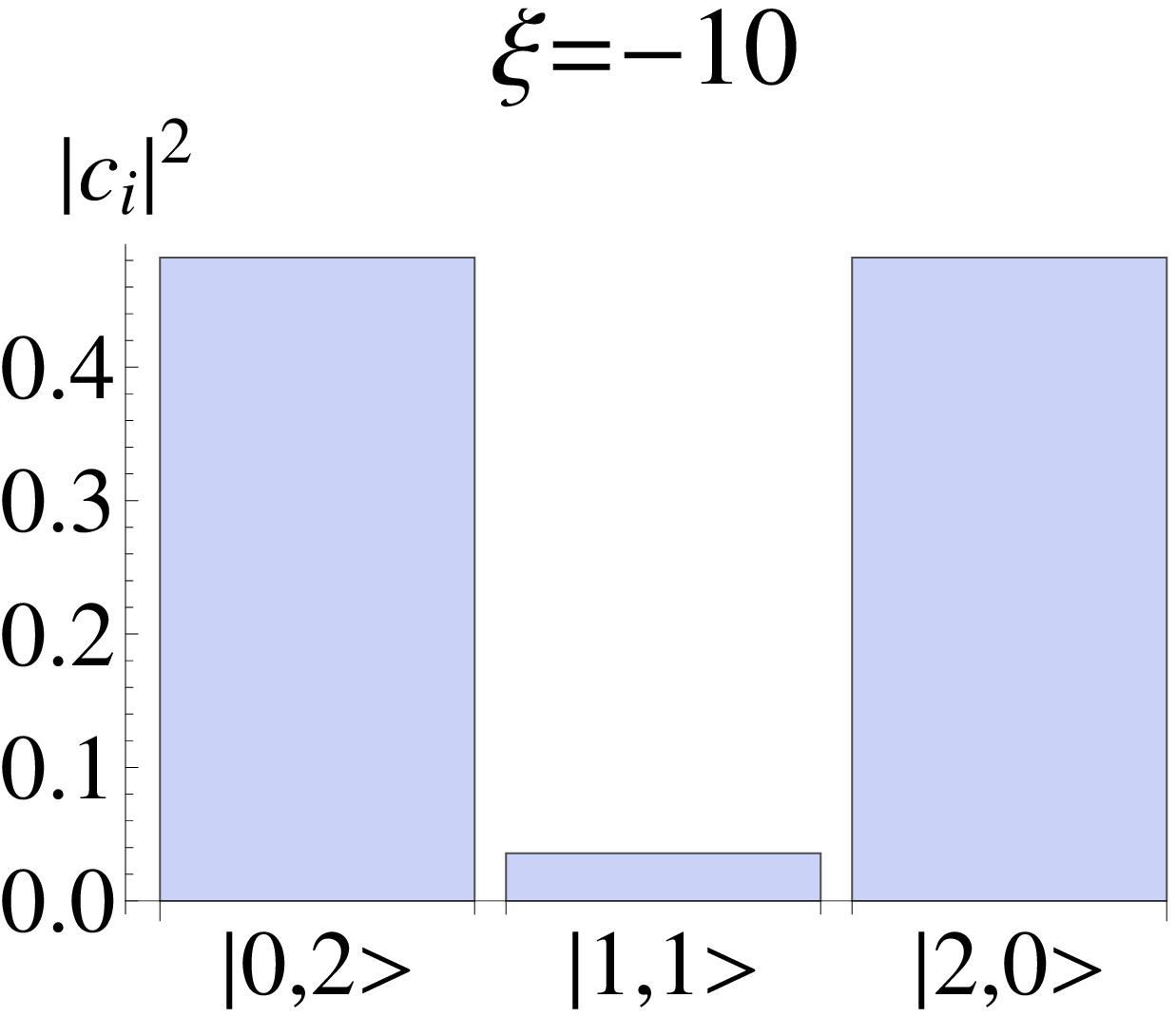}} \quad
{\includegraphics[width=.24\columnwidth]{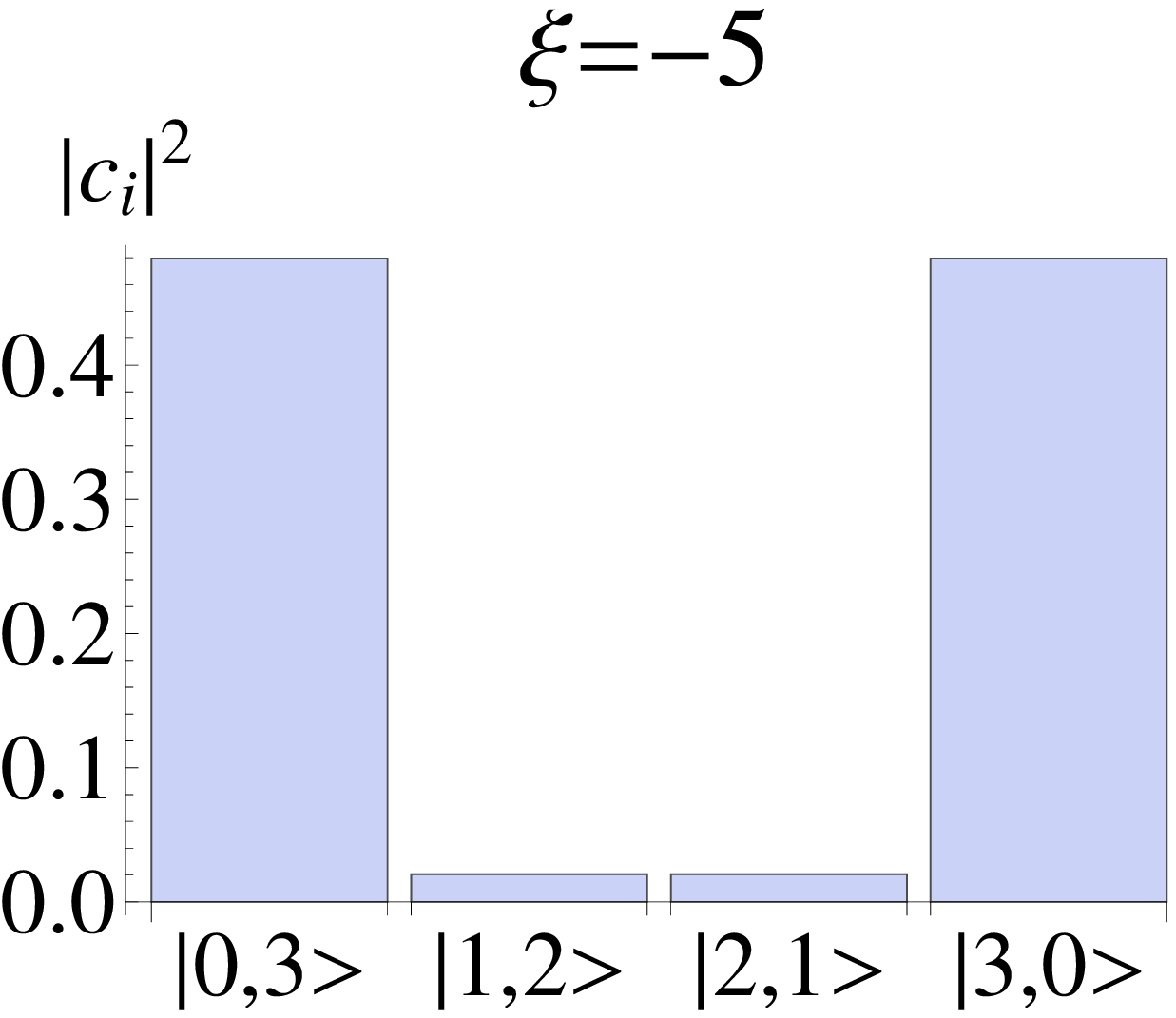}} \quad
{\includegraphics[width=.24\columnwidth]{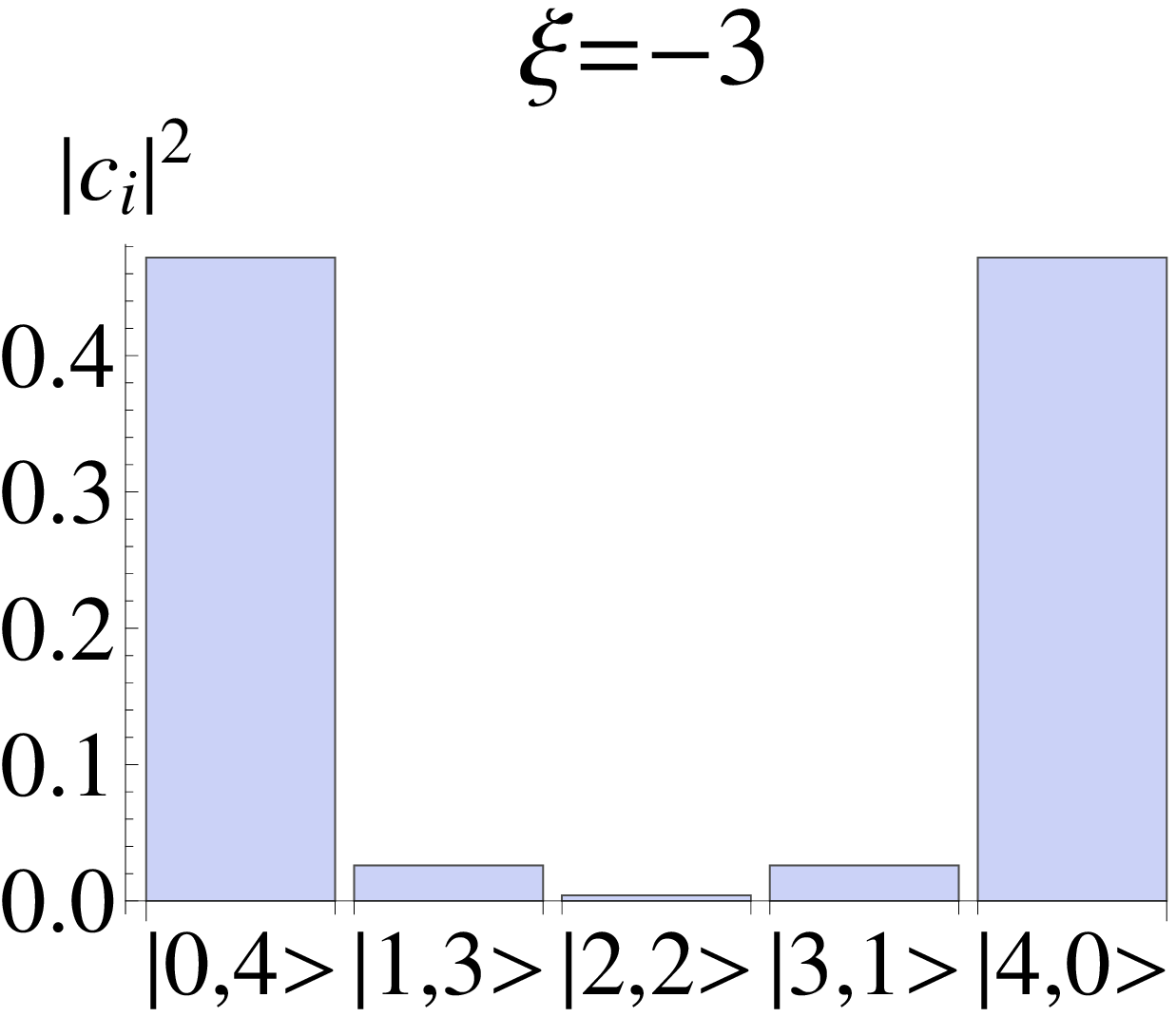}} \\
{\includegraphics[width=.24\columnwidth]{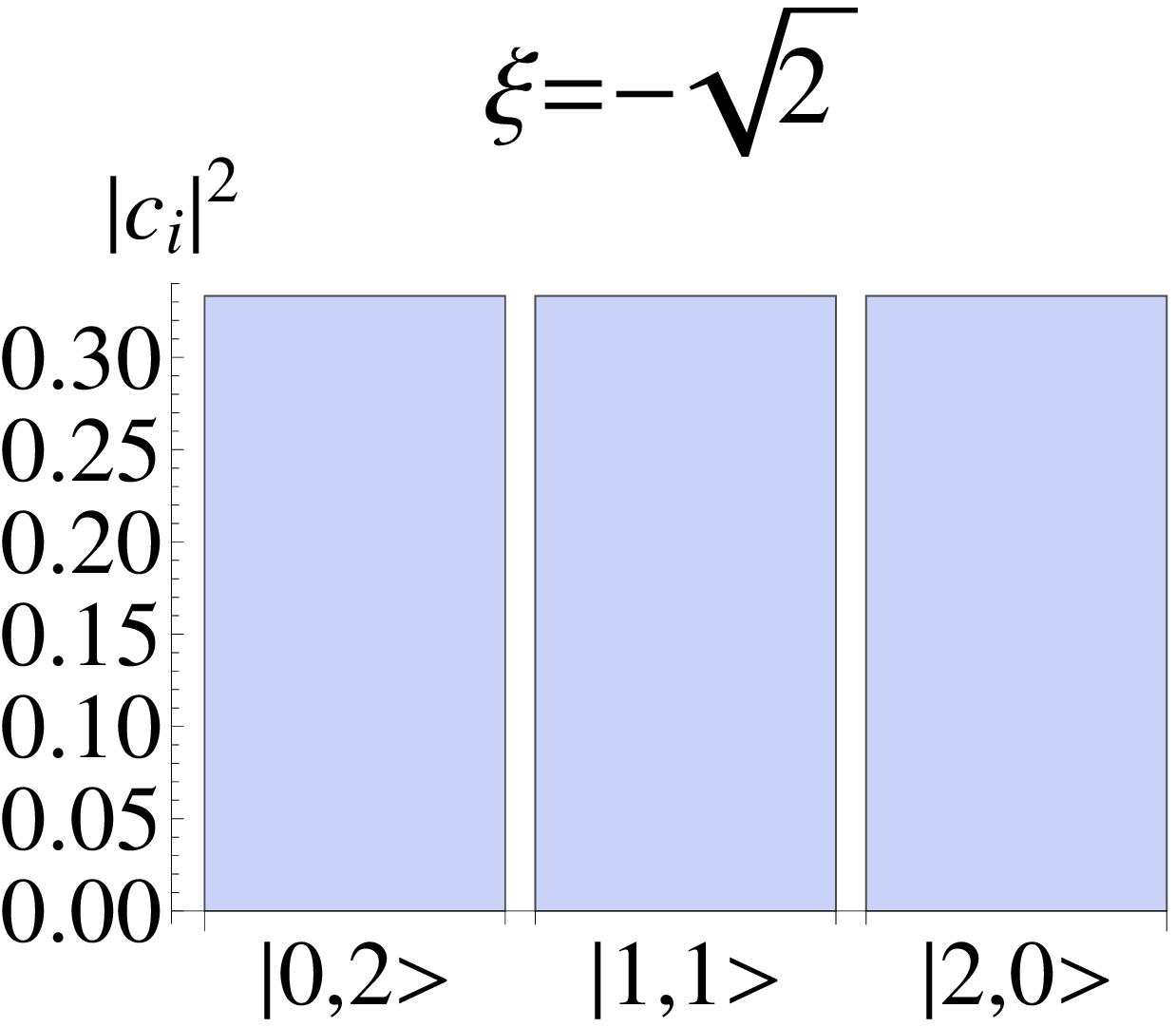}} \quad
{\includegraphics[width=.24\columnwidth]{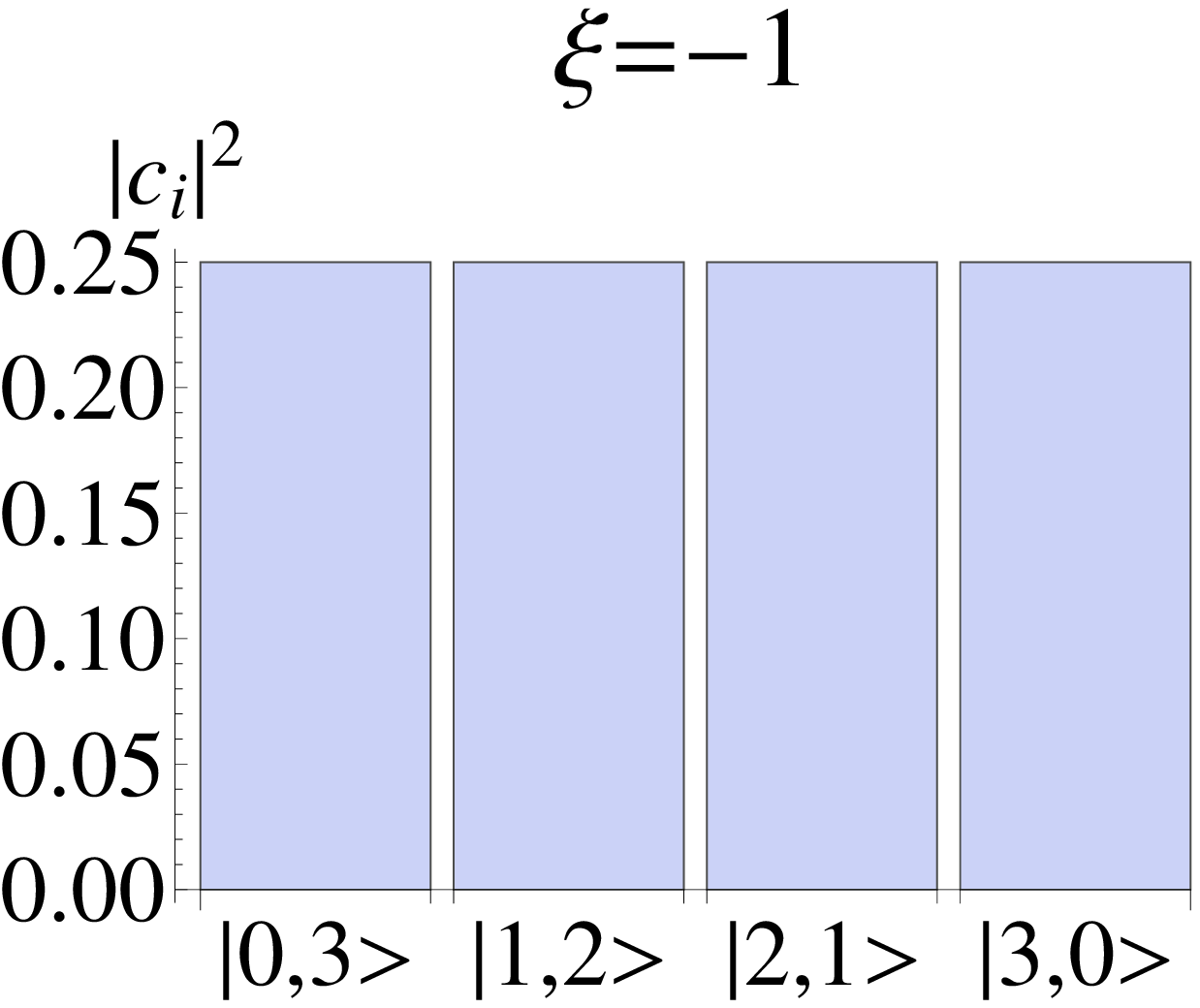}} \quad
{\includegraphics[width=.24\columnwidth]{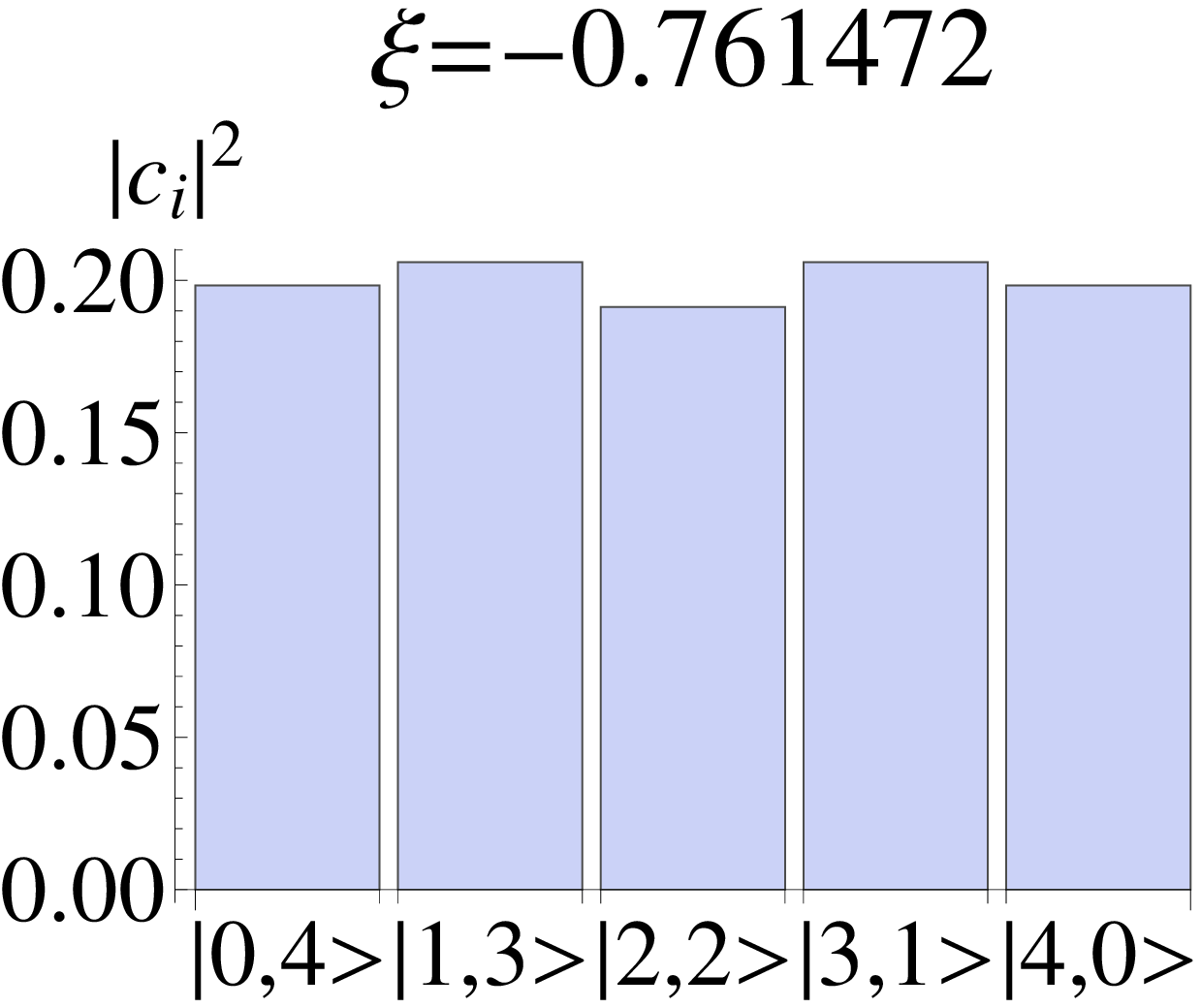}} \\
{\includegraphics[width=.24\columnwidth]{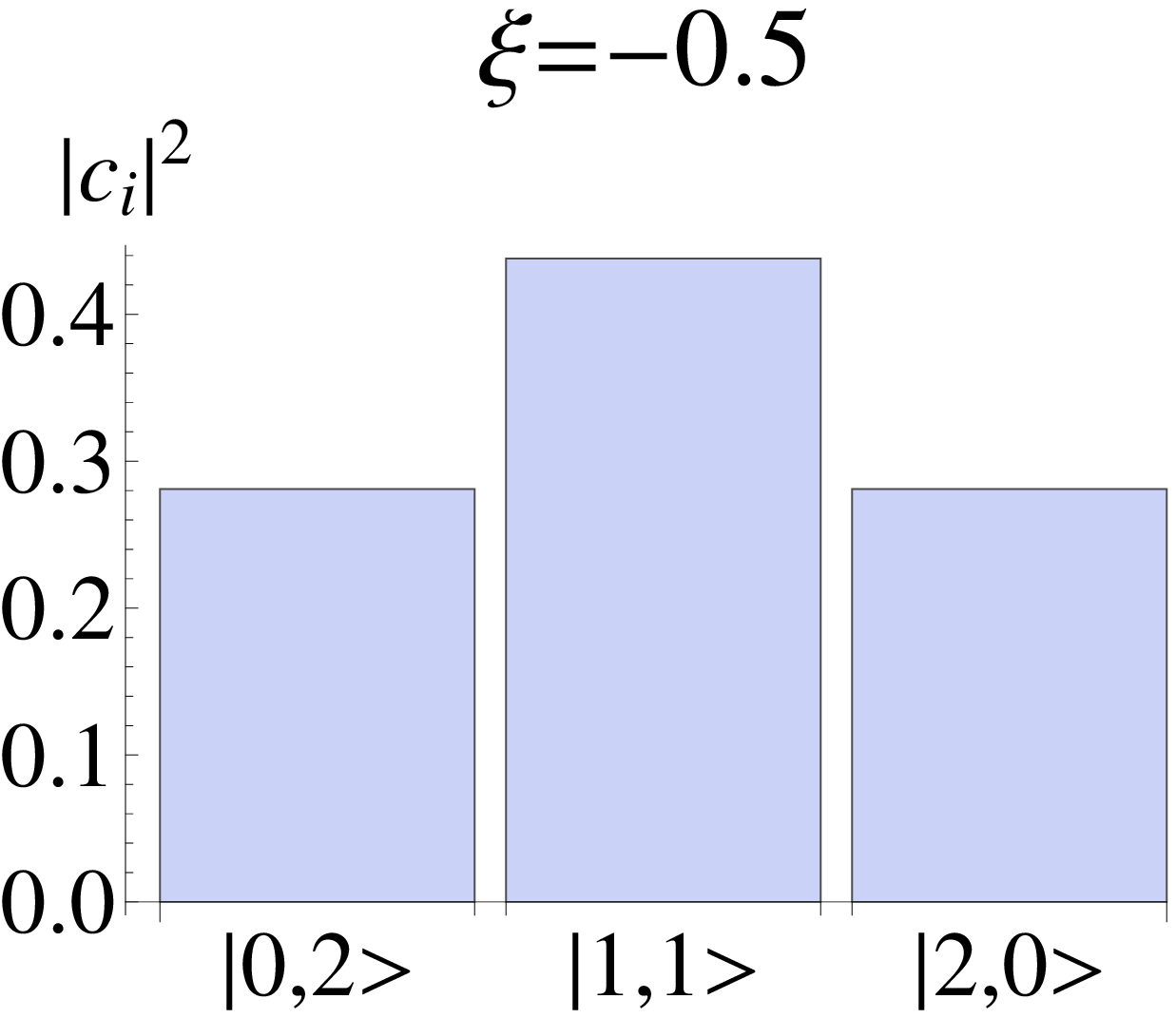}} \quad
{\includegraphics[width=.24\columnwidth]{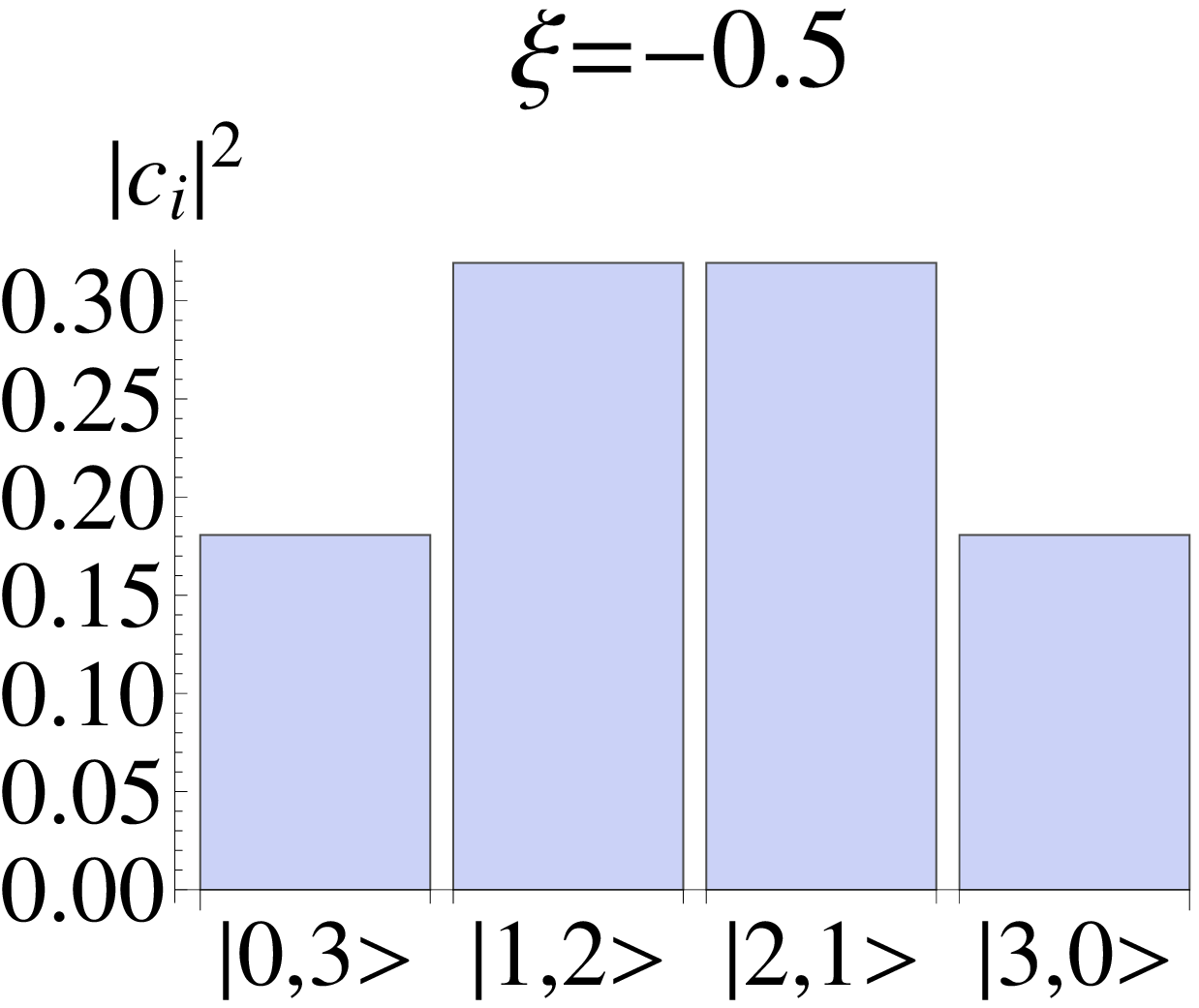}} \quad
{\includegraphics[width=.24\columnwidth]{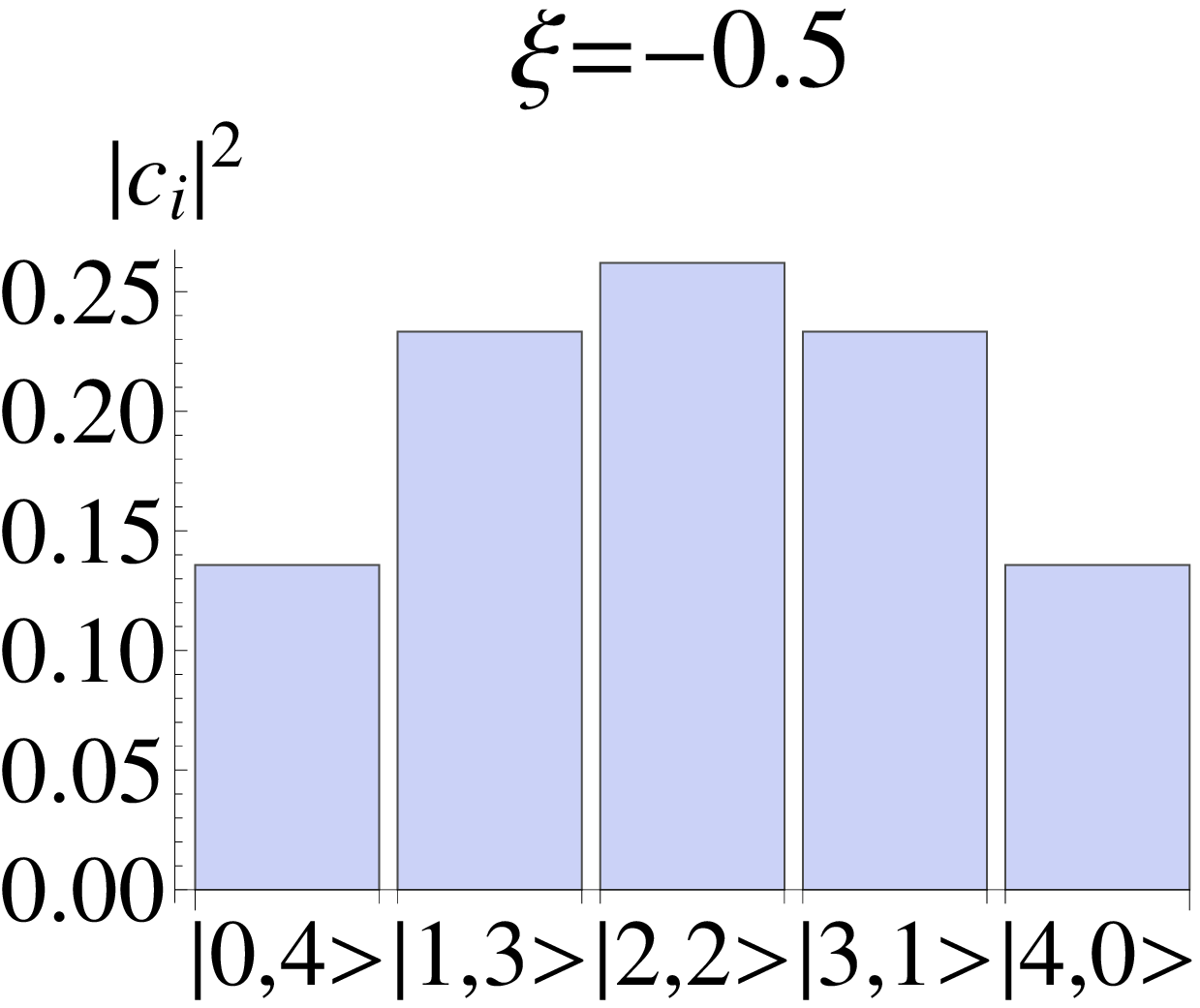}} \\
{\includegraphics[width=.24\columnwidth]{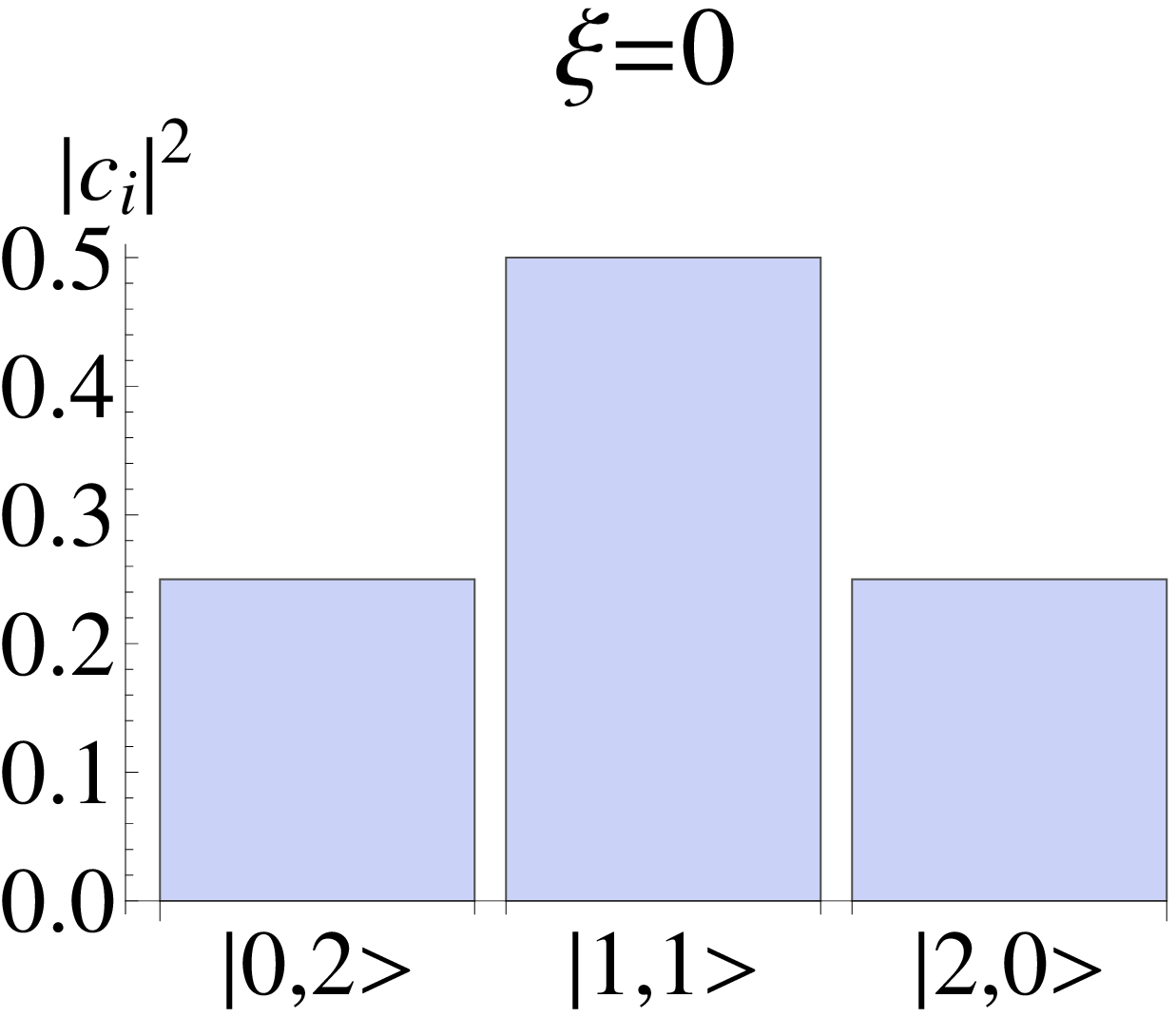}} \quad
{\includegraphics[width=.24\columnwidth]{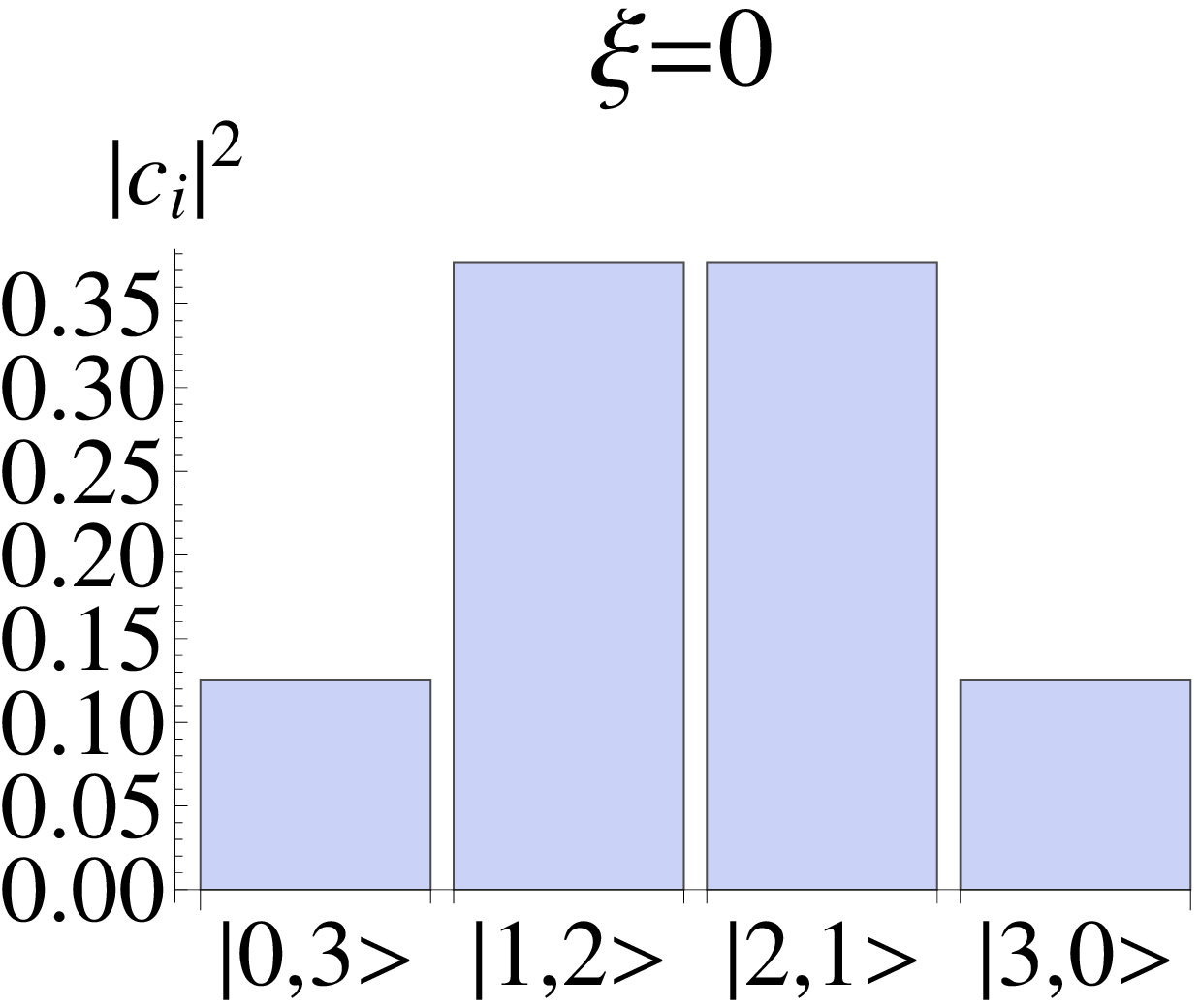}} \quad
{\includegraphics[width=.24\columnwidth]{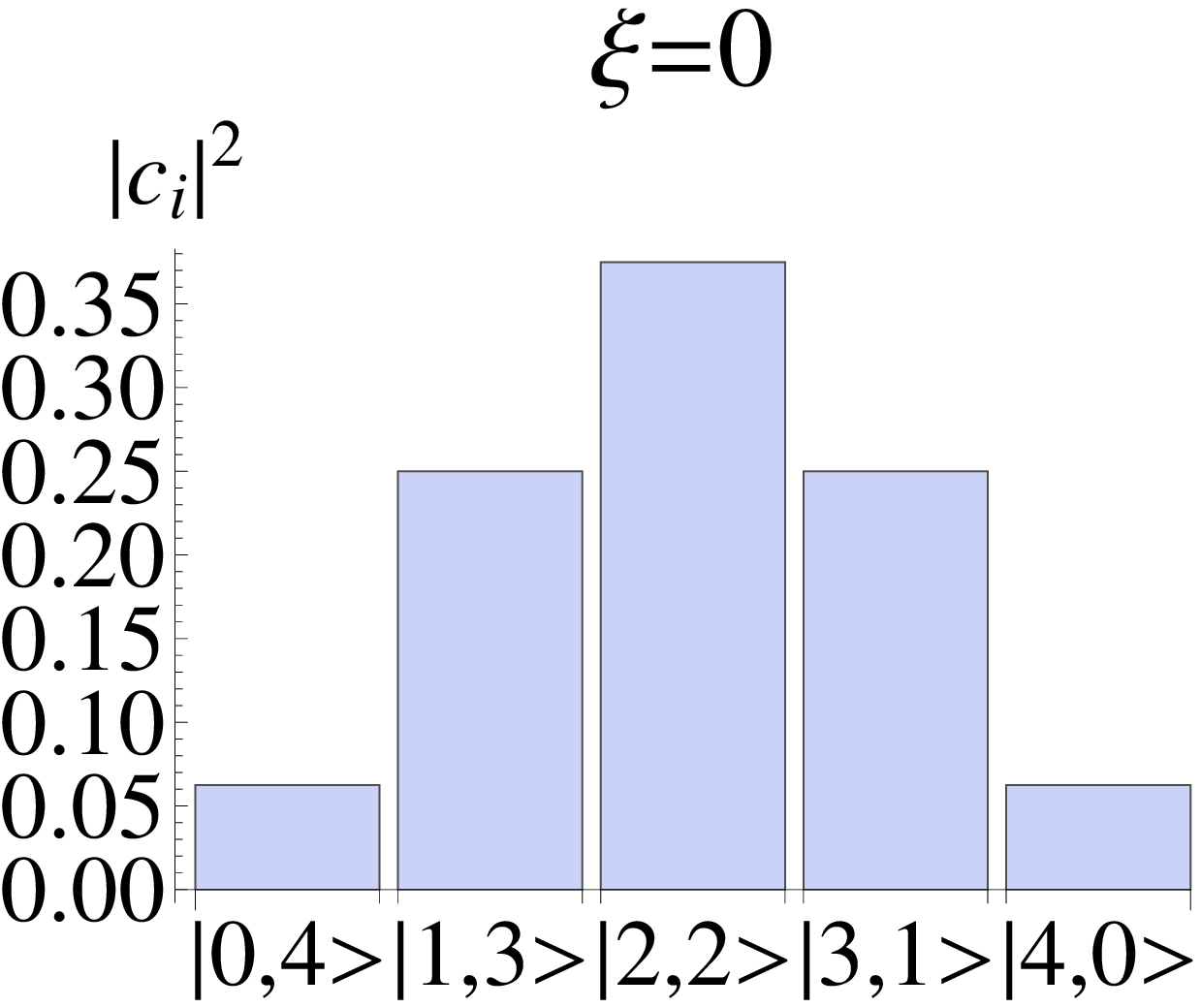}} \\
{\includegraphics[width=.24\columnwidth]{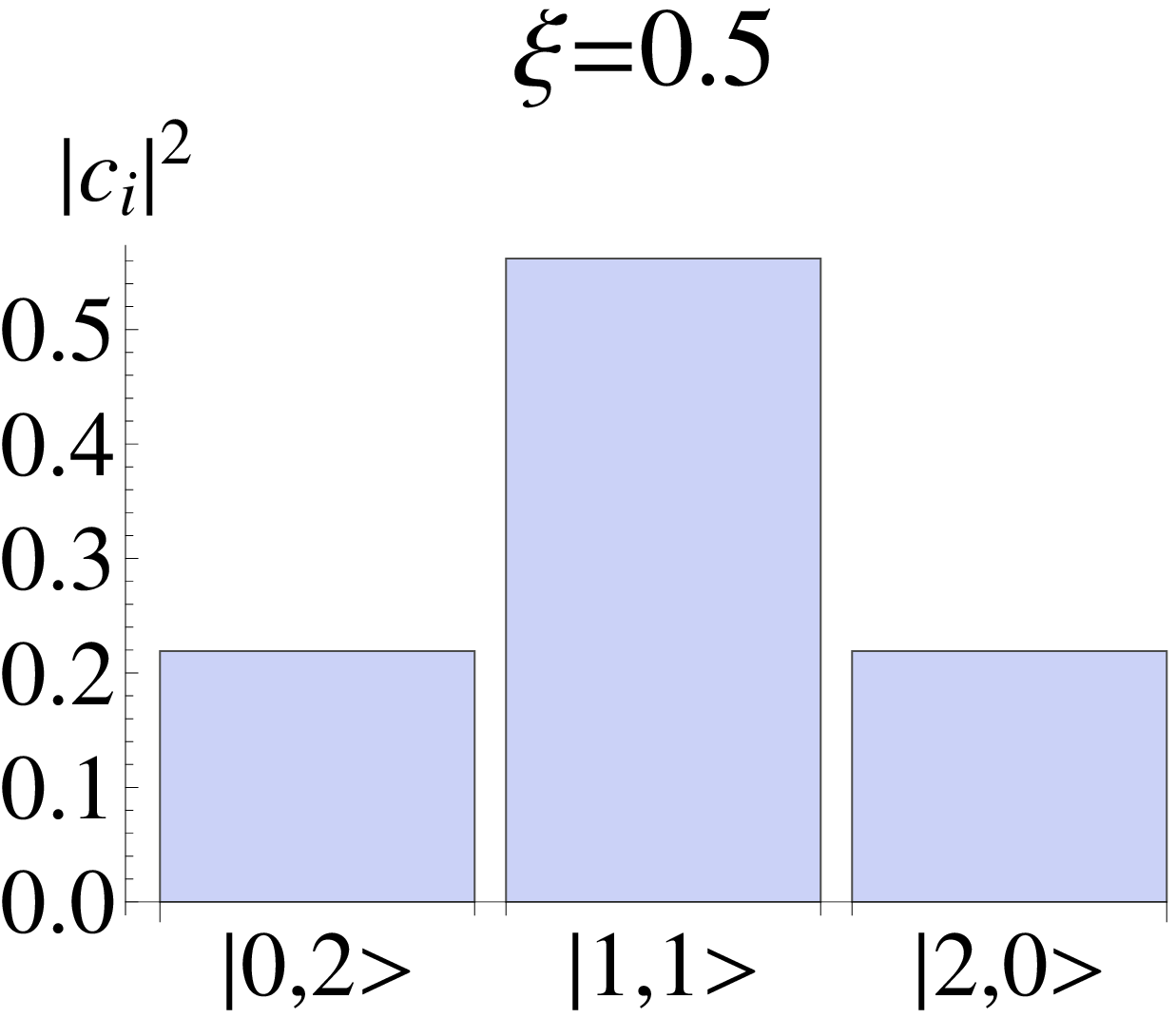}} \quad
{\includegraphics[width=.24\columnwidth]{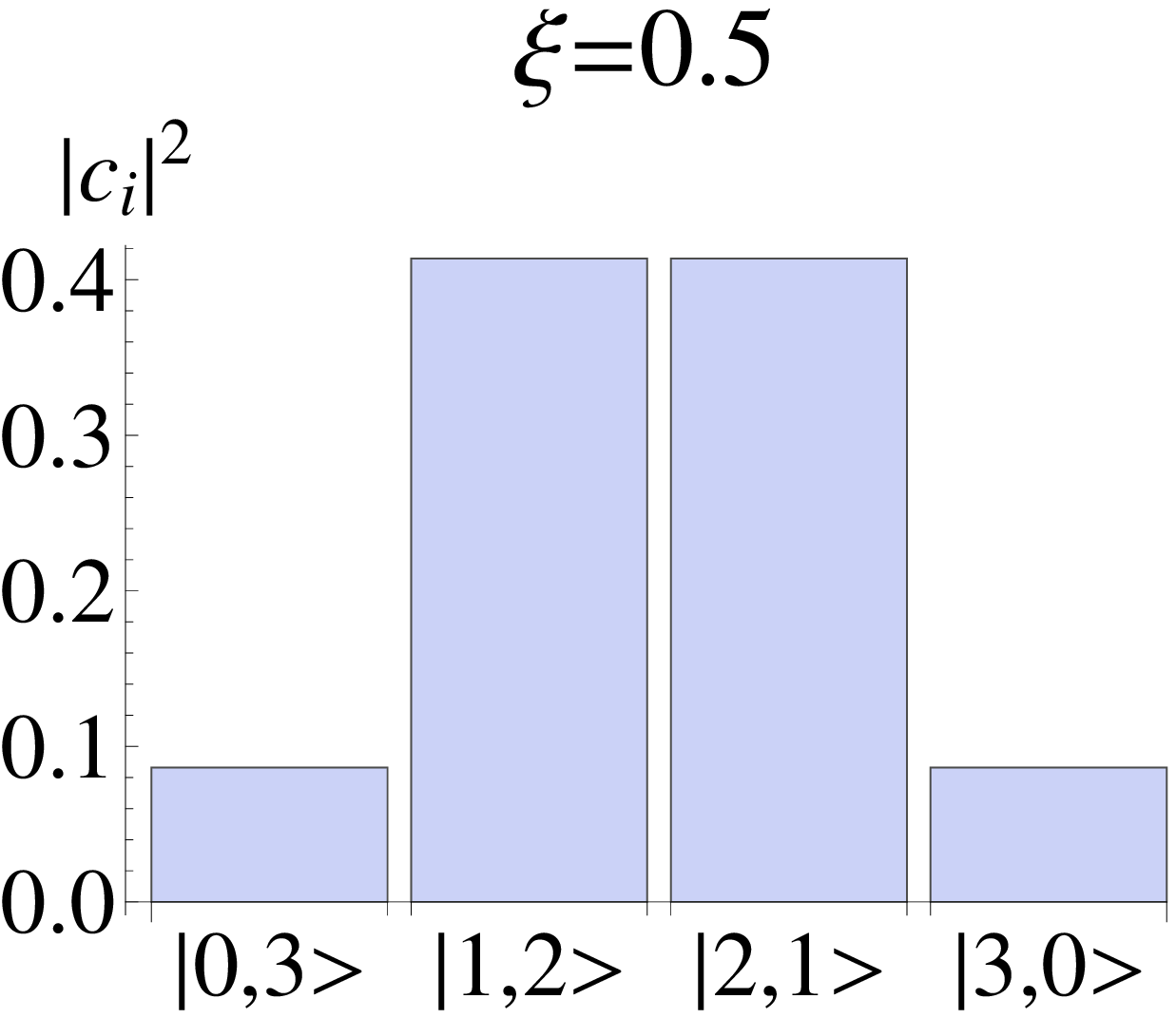}} \quad
{\includegraphics[width=.24\columnwidth]{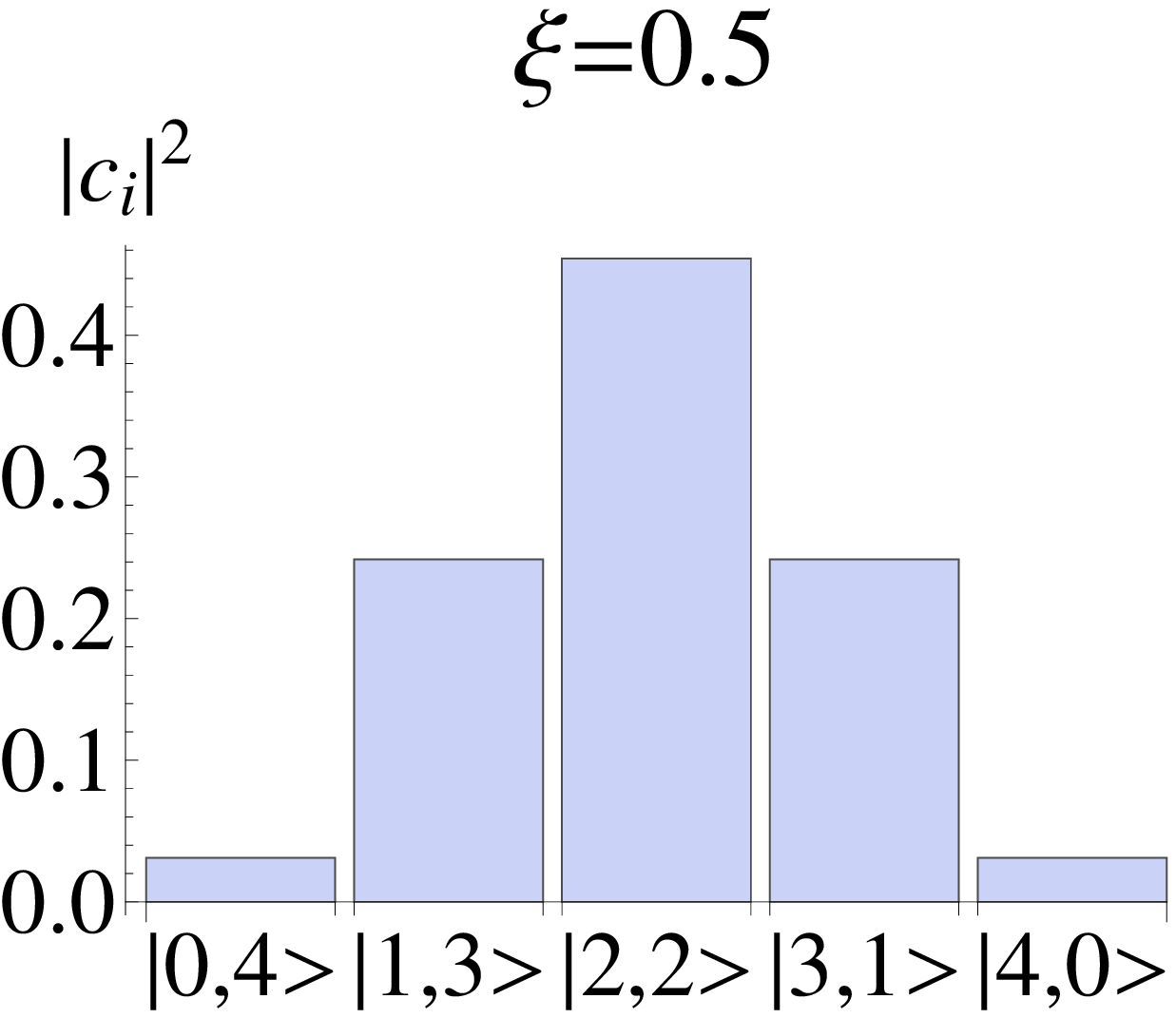}} \\
{\includegraphics[width=.24\columnwidth]{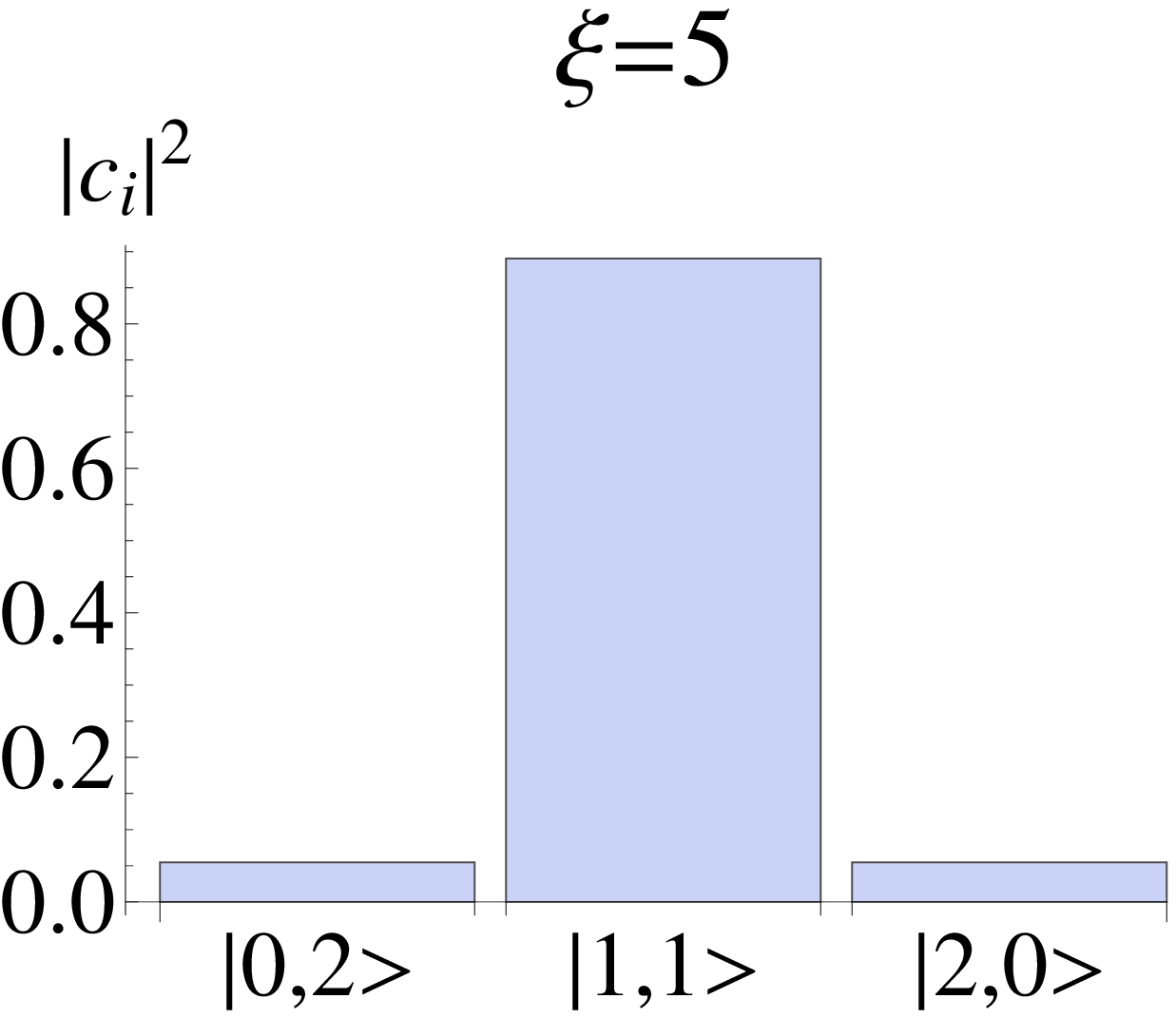}} \quad
{\includegraphics[width=.24\columnwidth]{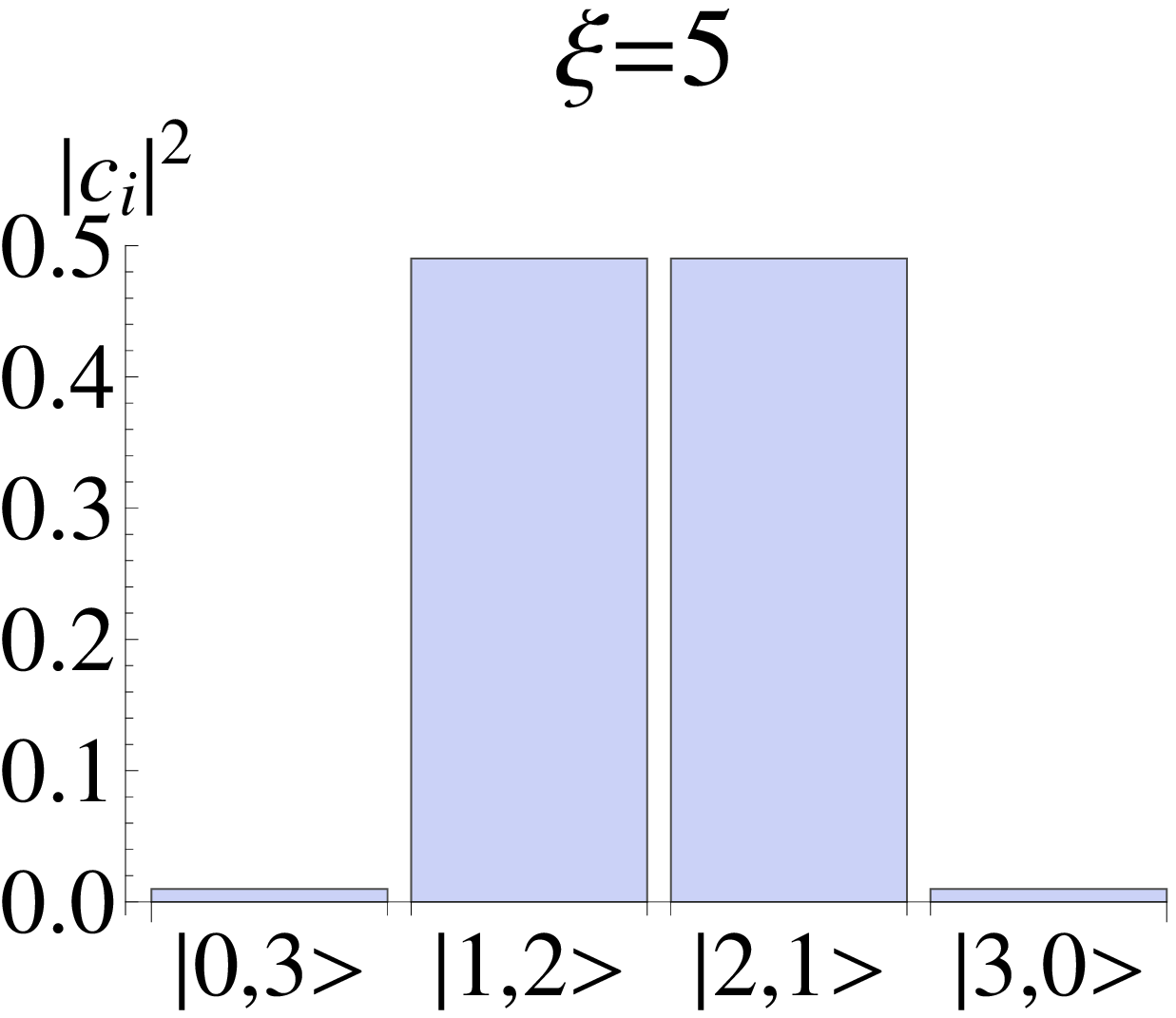}} \quad
{\includegraphics[width=.24\columnwidth]{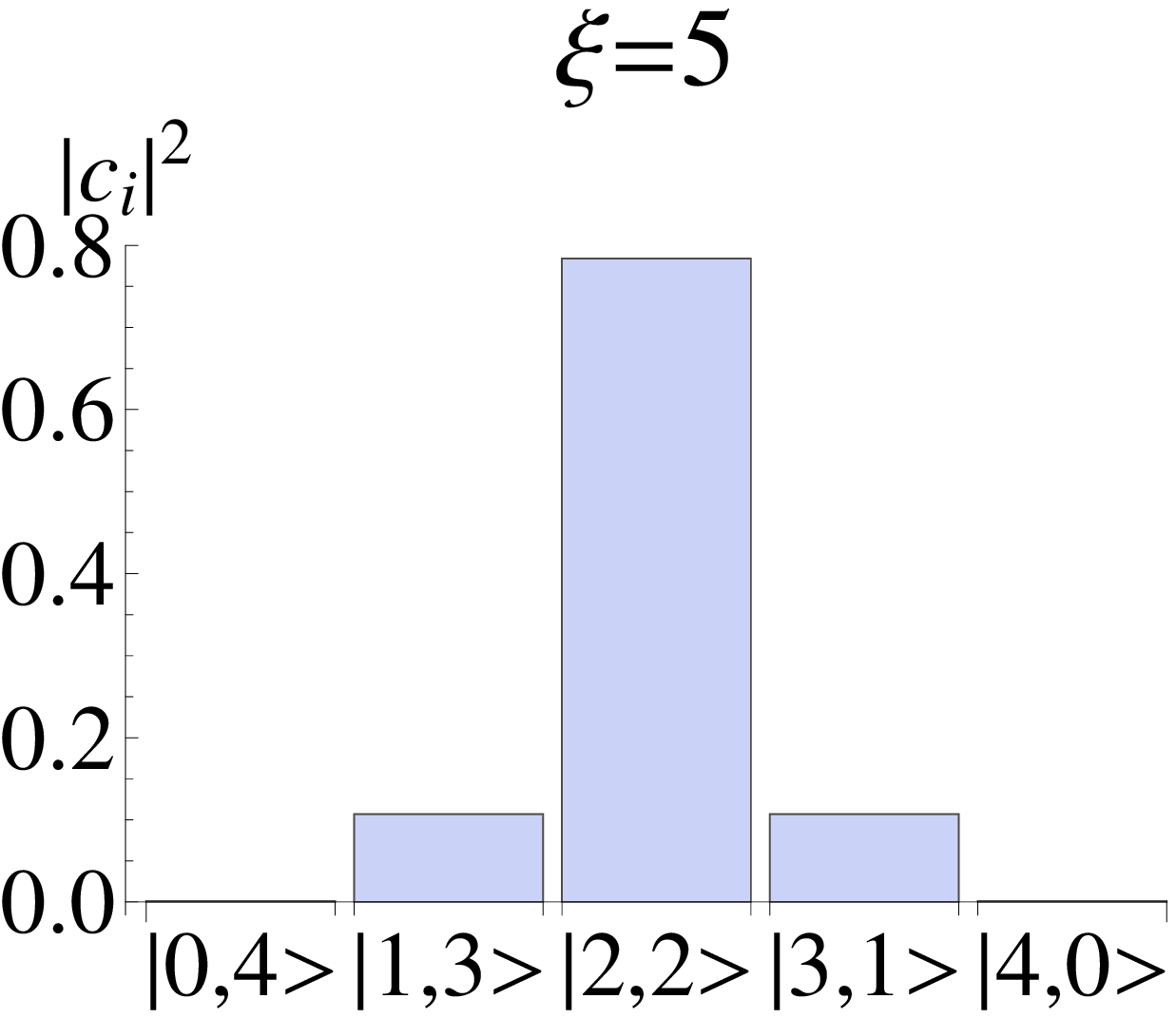}} \\
{\includegraphics[width=.24\columnwidth]{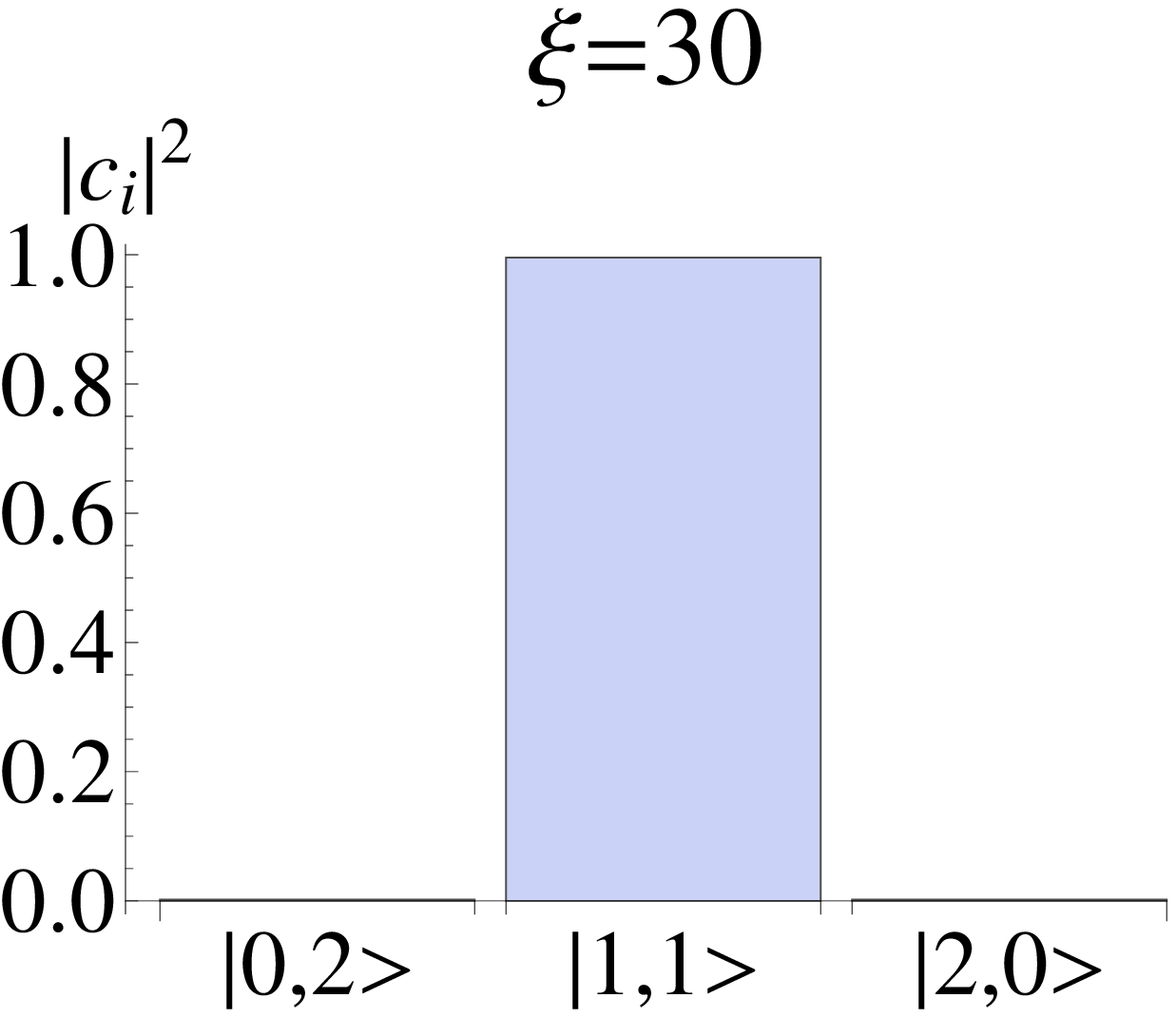}} \quad
{\includegraphics[width=.24\columnwidth]{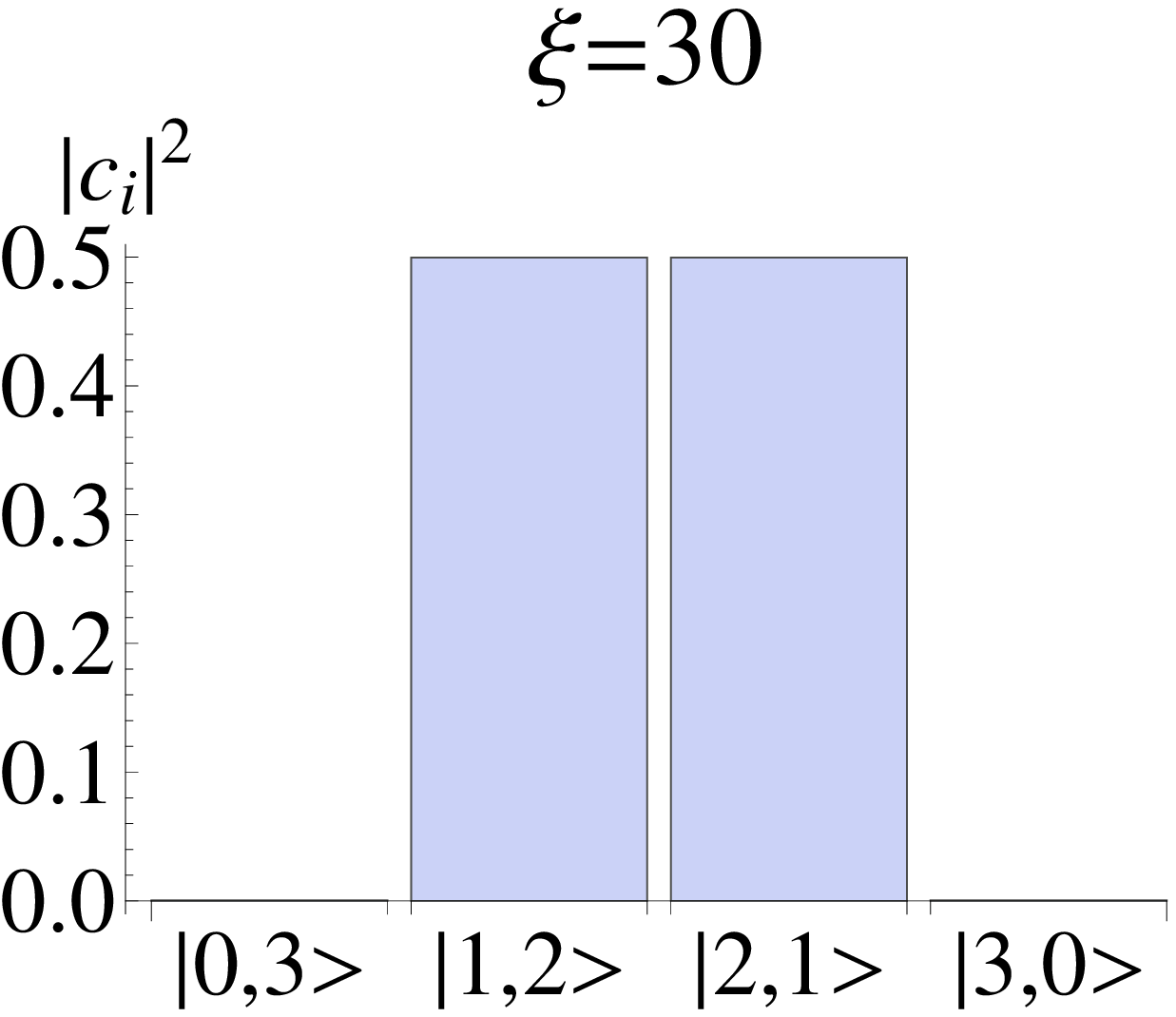}} \quad
{\includegraphics[width=.24\columnwidth]{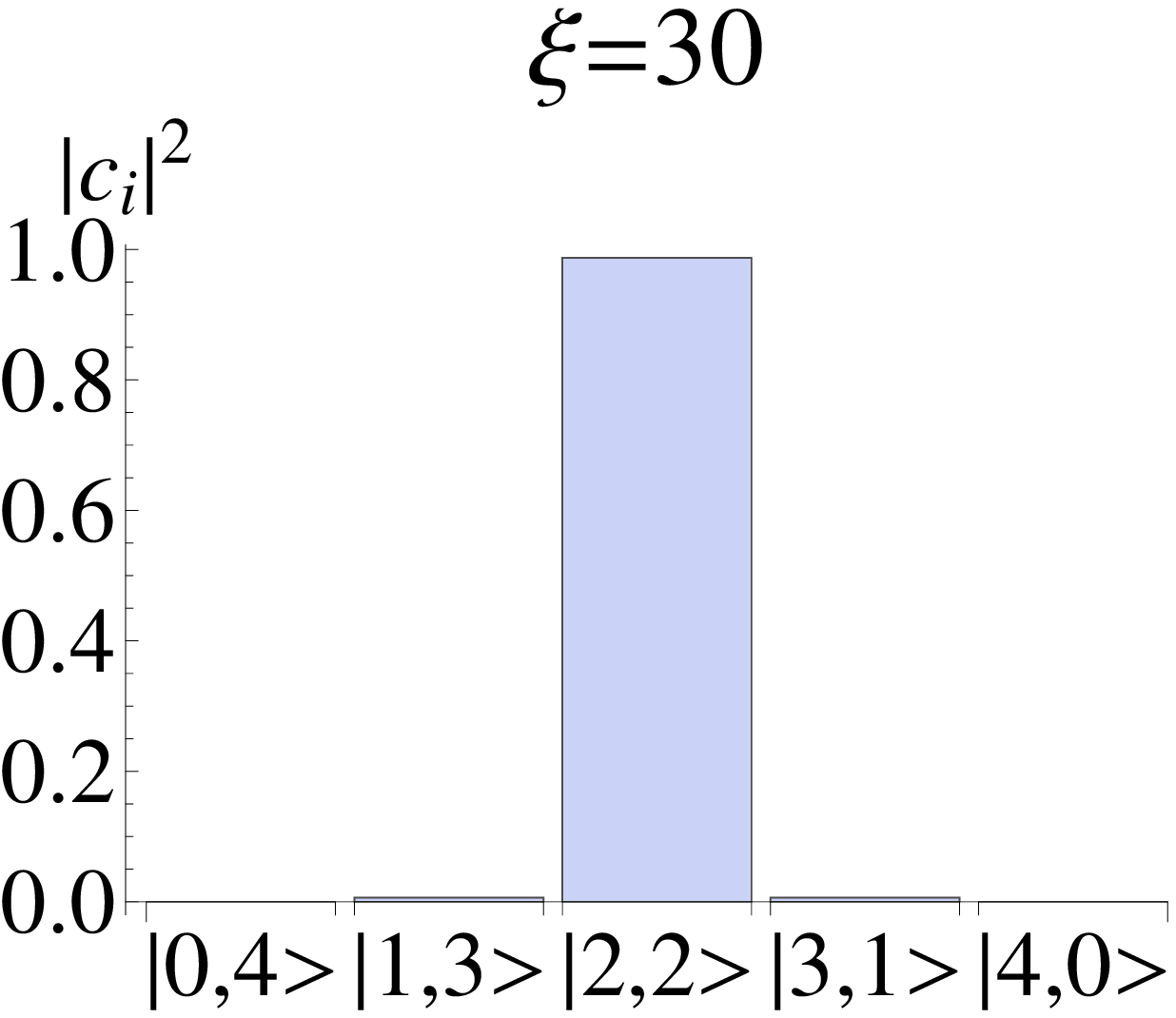}}
\caption{Vertical axis: probabilities $|c_i|^2$ for different values of $\xi=U/J$. Horizontal axis: kets $|i,N-i\rangle$ ($i=0,...,N$). Left: $N=2$. Middle: $N=3$. Right: $N=4$. At the third row (from the top): plots for $\xi$ signing the maximum of the entanglement entropies (\ref{s2}) ($N=2$), (\ref{s3}) ($N=3$), (\ref{s4}) ($N=4$). All the quantities are dimensionless.}
\label{result:coeff}
\end{figure}

In conclusions, we can say that a very strong boson-boson attraction tends to establish in the system a ground state given by a symmetric superposition of two fully populated Fock states both with $N=2,4$ and $N=3$ bosons. On the other hand, a very large interatomic repulsion induces different ground states depending if $N=2$ (separable twin-Fock state) or $N=3$ (symmetric superposition of quasi-fully populated Fock states).
%which describes a superfluid character through the presence of a fully-delocalized particle (hole) excitation).

On the repulsive side, the above described crossover, for even $N$, is reminescent of the quantum phase transition with optical-lattice-confined bosons theoretically predicted in \cite{jaksch} and experimentally observed by Greiner and co-workers \cite{greiner}. This transition - induced by varying the depth of the optical potential - is a transition from the superfluid phase (the hopping dominates the Hamiltonian, $J \gg U$. In this case each atom is spread out over the entire lattice) to the Mott insulator one (on-site interactions dominates the Hamiltonian, $U \gg J$. In this case, exact numbers of atoms are localized at individual lattice sites). Note that the even-odd difference (separable twin Fock state versus symmetric superposition of non-fully populated Fock states) which tends to become less relevant for larger particle numbers, indeed, is a well known Mott insulators feature, as commented, for example, in Ref. \cite{oguri}.
As commented before, we characterize the correlations of the ground state by calculating the Fisher information, the coherence visibility, and the entanglement entropy.
It is possible to achieve, in the case of $2$, $3$, and $4$ bosons, analytical formulas for these three parameters.

Let us start by  evaluating the Fisher information. To this end we employ at the right-hand side of Eq. (\ref {fi2}) the expressions for the coefficients $c_i$ given by Eq. (\ref{c2}) when $N=2$, by Eq. (\ref{c3}) when $N=3$, and by Eq. (\ref{c4}) when $N=4$ (with the normalization factors $A_2$ ($N=2$), $A_3$ ($N=3$), and $A_4$ ($N=4$) given by Eq. (\ref{a2}), Eq. (\ref{a3}), and Eq. (\ref{a4}) respectively; note that the $c_i$'s are real for any $N$, so that $c_{i}^*=c_{i}$). Then, for $N=2$ we obtain
\begin{equation}
\label{f2}
F = \frac{8}{16+\xi^2+\xi \sqrt{16+\xi^2}}
\;,\end{equation}
while for $N=3$ one gets
\begin{equation}
\label{f3}
F = \frac{27 + (1 + \xi + \sqrt{4 + \xi (2 + \xi)})^2}{9 (3 +
[1 + \xi + \sqrt{4 + \xi (2 + \xi)}]^2)}
\;.\end{equation}
For $N=4$, the Fisher information is given by
\begin{eqnarray}
\label{f4}
&&F=\frac{3\,\big(16+(\tilde E-6\xi)^2\big)}{e_4}\nonumber\\
&&e_4=64-(\tilde E-6 \xi)(54\xi^3-45\tilde E \xi^2+\nonumber\\
&&(\tilde E^2+4) (12\xi-\tilde E))
\;\end{eqnarray}
with $\tilde E$ being given by Eq. (\ref{egs4}).

We have studied the Fisher informations $F$ given by Eqs. (\ref{f2}), (\ref{f3}) and (\ref{f4}) as functions of the dimensionless parameter $\xi=U/J$, see the top panel of Fig. 2. As it can be seen from this figure, when the boson-boson interaction is strongly attractive ($\xi \ll -1$, this being correspondent to states close to the cat state (\ref{cat})) $F$ tends to $1$. In the deep repulsive regime, it can be observed that when $\xi \gg 1$ and $N=2,4$ (solid line, dot-dashed), when the ground state tends to a separable Fock state (\ref{fock}), $F$ tends to zero. With $N=3$ (dashed line), when the ground state tends to a superposition of two separable Fock states given by Eq. (\ref{pseudo:fock}), the Fisher information tends to a finite value (see also the Tabs. 1-3).

\begin{figure}[ht]
\epsfig{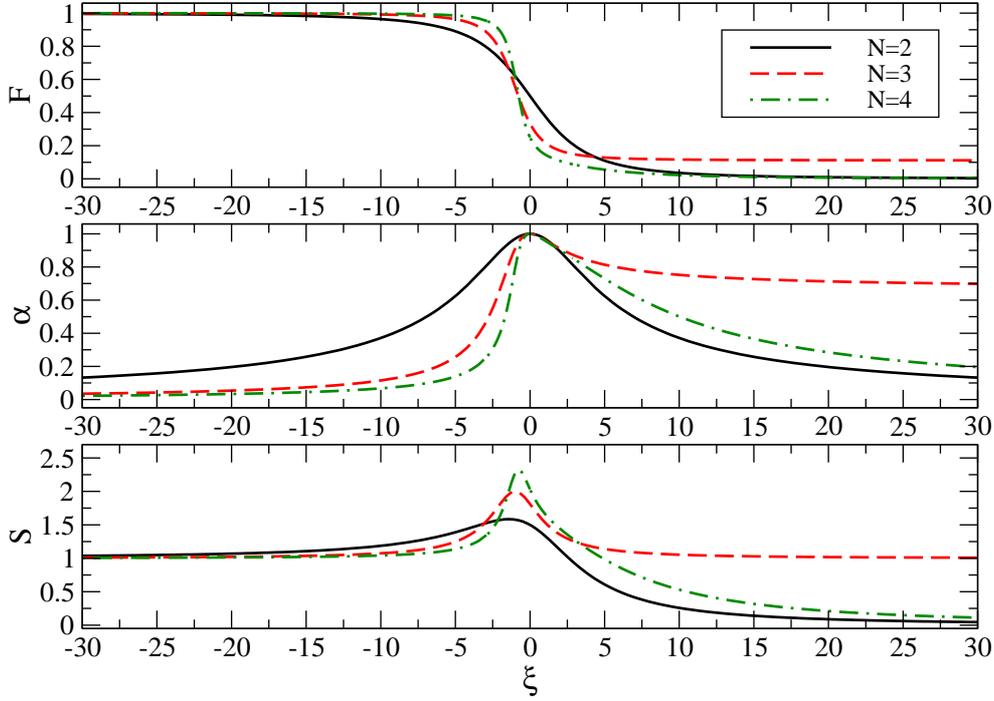}
\caption{(Color online). Fisher information $F$ (top panel), coherence visibility $\alpha$ (middle panel), entanglement entropy $S$ (bottom panel) vs scaled on-site interaction $\xi=U/J$. Solid line: $N=2$. Dashed line: $N=3$. Dot-dashed line: $N=4$. $F$, $\alpha$, $S$, and $\xi$ are dimensionless.}
\end{figure}

To obtain the coherence visibility $\alpha$, we use at the right-hand side of Eq. (\ref{visibility2}) the form of the $c_i$'s provided by Eq. (\ref{c2}) when $N=2$, by Eq. (\ref{c3}) when $N=3$, and by Eq. (\ref{c4}) when $N=4$ (with the normalization factors $A_2$ ($N=2$), $A_3$ ($N=3$), $A_4$ ($N=4$) given by Eq. (\ref{a2}), Eq. (\ref{a3}), Eq. (\ref{a4}), respectively). For $N=2$, we get
\begin{equation}
\label{alpha2}
\alpha = \frac{4 \left(\xi+\sqrt{16+\xi^2}\right)}{16+\xi^2+\xi \sqrt{16+\xi^2}}
\;,\end{equation}
for $N=3$
\begin{equation}
\label{alpha3}
\alpha = \frac{2(1 + \xi + \sqrt{4 + \xi (2 + \xi)}) (4 + \xi + \sqrt{4 + \xi (2 + \xi)})}{3 (3 +
  [1 + \xi + \sqrt{4 + \xi (2 + \xi)}]^2)}
\;,\end{equation}
and for $N=4$
\begin{equation}
\label{alpha4}
\alpha=\frac{6\sqrt{(\tilde E-6\xi)^2}(\sqrt{(18\xi^2-9\tilde E\xi +\tilde E^2-4)^2}+4)}{e_4}
\;\end{equation}
with $\tilde E$ given by Eq. (\ref{egs4}) and $e_4$ by the second row of Eq. (\ref{f4}).

We show the behavior of $\alpha$ (Eq. (\ref{alpha2}), Eq. (\ref{alpha3})), Eq. (\ref{alpha4})), when the scaled on-site interaction is varied, in the middle panel of Fig. 2. $\alpha$ reaches its maximum value ($\alpha=1$), both when $N=2,4$ (solid line, dot-dashed) and $N=3$ (dashed line), in the absence of boson-boson interaction that corresponds to the atomic coherent state (\ref{ACS}). For strongly attractive bosons, $\xi \ll -1$, the ground state is a cat-like state, and the coherence visibility approches to zero (see the solid (dot-dashed) line, $N=2(4)$, and the dashed one, $N=3$). When the repulsion between the bosons is sufficiently strong, $\xi \gg 1$, we can see that for $N=2,4$ (solid line, dot-dashed line)  - when the ground state tends to a fully incoherent state - $\alpha$ approaches to zero, while for $N=3$ (dashed line) - when the ground state is close to state (\ref{pseudo:fock}) - $\alpha$ is finite.

Finally, by employing the expressions of the coefficients $c_i$ provided by Eq. (\ref{c2}) in Eq. (\ref{ee}), we calculate the entanglement entropy $S$ for $N=2$, and get
\begin{equation}
\label{s2}
S=-A_{2}^{2}\bigg(2\log_2[2A_2^{2}] +\frac{s_2}{4}\log_2[\frac{s_2A_2^{2}}{4}]\bigg)
\;,\end{equation}
where $A_2$ is the normalization factor given by Eq. (\ref{a2}) and
\begin{equation}
\label{s2constant}
s_2=\frac{(\xi+\sqrt{16+\xi^2})^2}{2}
\;.\end{equation}

By employing the expressions of the coefficients $c_i$ provided by Eq. (\ref{c3}) in Eq. (\ref{ee}), we calculate the entanglement entropy $S$ for $N=3$, which has the following expression

\begin{equation}
\label{s3}
S=-\frac{2A_{3}^2}{3}\bigg(s_{3}^2\,\log_2[\frac{s_3}{4\sqrt{4+\xi(\xi+2)}}]+3\log_2[A_{3}^2]\bigg)
\;,\end{equation}
where $A_3$ is the normalization factor given by Eq. (\ref{a3}) and
\begin{equation}
\label{s3constant}
s_3=(1+\xi+\sqrt{4+\xi(\xi+2)})
\;.\end{equation}

By following the same path for $N=4$ (i.e. by using formulas given by Eq. (\ref{c4}) in Eq. (\ref{ee})), one obtains the following entanglement entropy
%\begin{widetext}
\begin{eqnarray}
\label{s4}
&&S=\log_2 d_4-d_{4}^{-1}\bigg(48 \log_2 24+24(6\xi-\tilde E)^2\log_2[6(6\xi-\tilde E)]\nonumber\\
&&+2(18\xi^2-9\tilde E\xi+\tilde E^2-4)^2\log_2[(18\xi^2-9\tilde E\xi+\tilde E^2-4)]\bigg)\;,\nonumber\\
\end{eqnarray}
%\end{widetext}
where $\tilde E$ is given by Eq. (\ref{egs4}) and $d_4$ by the second row of Eq. (\ref{a4}).

Note that the results stated for $N=2$ by Eqs. (\ref{f2}) (Fisher information), (\ref{alpha2}) (coherence visibility), and (\ref{s2}) (entanglement entropy) coincide with those that we obtained in \cite{main}.

The bottom panel of Fig. 2 shows the entanglement entropies (\ref{s2}), (\ref{s3}), and (\ref{s4}) as functions of the scaled on-site interaction $\xi=U/J$. In the limit $\xi \rightarrow -\infty$, corresponding to the emergence of the cat-like state (\ref{cat}), $S$ tends to one both when $N=2,4$ (solid line, dot-dashed line) and $N=3$ (dashed line). By analyzing the plot of $S$ versus $\xi$ it is possible to observe that - as discussed in \cite{main} - the greater is the number of bosons the closest to zero is the negative $\xi$ (which, in fact, is equal to $-\sqrt{2}\simeq -1.41421$ when $N=2$, $-1$ when $N=3$, and is $\simeq -0.761472$ when $N=4$) for which $S$ attains its maximum value.
For $N=2$ and $N=3$, the latter value coincide with that predicted theoretically, i.e. $\log_2(N+1)$ (see the comment at the end of the previous section): in the ground state of the two-site Bose-Hubbard Hamiltonian all the Fock states $|i,N-i\rangle$ have the same probability $|c_i|^2=1/(N+1)$ for any $i$, as one can see from the left and the middle panels of the third row (from the top) of Fig. 1. When $N=4$, instead, the maximum of $S$ does not coincide with $\log_2(N+1)$, as it can be observed from the right plot of the third row (from the top) of Fig.1, where the $|c_i|^2$, corresponding to the interaction signing the maximum of $S$ (\ref{s4}), are different from each other.

As conclusive remarks, we note that when $N=3$, $S$ approaches to $1$ both in the limit $\xi \rightarrow -\infty$ and in the limit $\xi \rightarrow +\infty$, as one can observe from the dashed line in the bottom panel of Fig. 2.
Moreover, we observe that the plot reported in the bottom panel of Fig. 2 shows that the cat-like state (\ref{cat}) ($\xi \ll -1$, deeply attractive bosons) is not the maximally entangled ground state achievable in our system.

%In order to show that the results about the limit states we obtained are valid also for larger numbers of particles, we evaluated $F$, $\alpha$ and $S$ for $N=10$ and $N=15$
%(we used small variations of the C++ code in appendix). It is easy to find that the functions tend to the values listed in table \ref{parameters:limits},
%while their variation depending on $\xi=U/J$ is represented in fig.~\ref{fas}.

A remarkable point emerges from our analysis. The three ground-state characterizing parameters ($F$, $\alpha$, $S$) in the deep repulsive regime exhibit very different behavior depending on if $N=2,4$ or $N=3$, as it can be seen from Fig. 2 and from Tabs. 1-3. In fact, when $\xi=U/J\rightarrow +\infty$, the Fisher information, the coherence visibility, and entanglement entropy are all equal to zero if the ground state is the state (\ref{fock}) ($N=2,4$); they are finite, instead, if the ground state is the state (\ref{pseudo:fock}) ($N=3$). Note that this circumstance is quite general. In fact, the states (\ref{fock}) and (\ref{pseudo:fock}) are the ground states of the two-site Bose-Hubbard Hamiltonian in the limit $\xi \rightarrow +\infty$ for any even $N$ and any odd $N$, respectively. In particular, we want to stress that when the boson-boson interaction is strongly repulsive, the ground state of the two-site BH Hamiltonian is not quantum entangled when $N$ is even, while it is a quantum entangled state when $N$ is odd.

\subsection{Fisher information and coherence visibility via the Hellmann-Feynmann theorem}

At this point, it is interesting to observe that it is possible to achieve the above formulas for $F$ - Eqs. (\ref{f2}),(\ref{f3}),(\ref{f4}) -  and $\alpha$ - Eqs. (\ref{alpha2}),(\ref{alpha3}), (\ref{alpha4})  - also by exploiting the Hellmann-Feynman theorem (HFT) \cite{cohen}.

By using this theorem a relation can be established between the Fisher information $F$ and
the first partial derivative of the ground-state energy $E$ with respect to $U$, $\displaystyle{\frac{\partial E}{\partial U}}$. According to the HFT,
we have that \cite{cohen}
\begin{equation}
\label{hft1}
\frac{\partial E}{\partial U}=\langle E|\frac{\partial \hat{H}}
{\partial U}|E\rangle
\;.\end{equation}
If the  properties $\langle E |\hat{n}_L|E\rangle=\langle  E|\hat{n}_R|E\rangle=N/2$ (Eq. (\ref{lrs})) and $\langle E |\hat{n}_L|E\rangle+\langle  E|\hat{n}_R|E\rangle=N$ are used in Eq. (\ref{hft1}), we get
\begin{equation}
\label{ezerou1}
\frac{\partial E}{\partial U}=\langle E| \hat{n}^{2}_L |E\rangle-\frac{N}{2}
\;.\end{equation}
On the other hand, again thanks to $\langle E |\hat{n}_L|E\rangle=\langle  E|\hat{n}_R|E\rangle=N/2$ and  $\langle E |\hat{n}_L|E\rangle+\langle  E|\hat{n}_R|E\rangle=N$ ,  Eqs. (\ref{qfi}) and (\ref{fi}) give rise to
\begin{equation}
\label{ezerou2}
F=\frac{4}{N^2}\langle E | \hat{n}^{2}_L|E\rangle-1
\;,\end{equation}
so that (as also commented in \cite{main})
\begin{equation}
\label{fihf}
F=\frac{4}{N^{2}}\frac{\partial E}{\partial U}+\frac{2}{N}-1
\;.\end{equation}
We have therefore to know the energy of the ground state. When $N=2$ this energy is given by Eq. (\ref{egs2}), when $N=3$ by Eq. (\ref{egs3}), and when $N=4$ by Eq. (\ref{egs4}). By keeping in mind that in these three latter equations $\xi=U/J$ and the energies are measured in units of $J$, one can resort from $\tilde E$ to $E$ and obtain Eq. (\ref{f2}) for $N=2$, Eq. (\ref{f3}) for $N=3$, and Eq. (\ref{f4}) for $N=4$.

The Hellmann-Feynman theorem provides a relation between the coherence visibility $\alpha$
and the first partial derivative of the ground-state energy $E$ with respect to $J$ as well. Let us start from the fact that, according to the HFT,
one has that \cite{cohen}
\begin{equation}
\frac{\partial E}{\partial J}=\langle E|
\frac{\partial \hat{H}}{\partial J}|E\rangle
\;.\end{equation}
By using the fact that the coefficients $c_i$ involved in the expansion of $|E\rangle$ ($N=2,3,4$) are real (see Eqs. (\ref{c2}), (\ref{c3}), (\ref{c4})  jointly to Eqs. (\ref{a2}),(\ref{a3}),(\ref{a4})), we can write that
$\langle E| \hat{a}^\dagger_L\hat{a}_R|E\rangle=\langle E| \hat{a}^\dagger_R\hat{a}_L|E\rangle$, and then
\begin{equation}
\label{ezeroj1}
\frac{\partial E}{\partial J}=-2\langle E| \hat{a}^\dagger_L\hat{a}_R|E\rangle
\;.\end{equation}
On the other hand, again in force to $\langle E| \hat{a}^\dagger_L\hat{a}_R|E\rangle=\langle E| \hat{a}^\dagger_R\hat{a}_L|E\rangle$, we have that the coherence visibility (\ref{visibility}) is
\begin{equation}
\label{ezeroj2}
\alpha=\frac{2 \langle E| \hat{a}^\dagger_L\hat{a}_R|E\rangle}{N}
\end{equation}
so that (see also \cite{main})
\begin{equation}
\label{alphahf}
\alpha =-\frac{1}{N}\frac{\partial E}{\partial J}\;.
\end{equation}
Also in this case, we use in Eq. (\ref{alphahf}) the results given by Eq. (\ref{egs2}), $N=2$, by Eq. (\ref{egs3}), $N=3$, and by Eq. (\ref{egs4}) (by keeping in mind that $\xi=U/J$ and the energies are measured in units of $J$) and get Eq. (\ref{alpha2}) for $N=2$, Eq. (\ref{alpha3}) for $N=3$, and Eq. (\ref{alpha4}) for $N=4$.\\

\section{Conclusions}

We have investigated a finite number $N$ of (both attractively and repulsively) interacting bosonic atoms confined
by a one-dimensional double-well shaped potential. Within the two-site Bose-Hubbard model framework, we have carried out the zero-temperature analysis for $N=2$, $N=3$, and $N=4$ bosons by finding analytical formulas for the eigenvectors and eigenvalues of the corresponding ground states. These have been characterized by analytically calculating the Fisher information, the coherence visibility, and the entanglement entropy. We have studied these parameters by varying the boson-boson interaction strength (ranging from strong attractions to strong repulsions) which is the key quantity in determining the kind of ground state sustained by the two-site Bose-Hubbard Hamiltonian. In particular, we have commented on the difference, existing in the deep repulsive regime, between the structure of the ground state in the presence of an even number of bosons and that with an odd number of particles. We have pointed out, in particular, that the ground state of the two-site Bose-Hubbard Hamiltonian is not quantum entangled when $N$ is even, while it is a quantum entangled state when $N$ is odd.\\

\begin{acknowledgement}
This work has been supported by MIUR (PRIN 2010LLKJBX). GM and LS acknowledge financial support from
the University of Padova (Progetto di Ateneo 2011) and Cariparo Foundation (Progetto di Eccellenza 2011). GM acknowledges financial support also from Progetto Giovani 2011
of University of Padova.
\end{acknowledgement}

% --------------------------------------------------------------


\begin{thebibliography}{10}

\bibitem{oliver} O. Morsch and M. Oberthaler,
Rev. Mod. Phys. {\bf 78}, 179 (2006).

\bibitem{smerzi} S.Raghavan, A. Smerzi, S. Fantoni, R. Shenoy,
Phys. Rev. A {\bf 59}, 620 (1999).

\bibitem{stringari} L. Pitaevskii and S. Stringari,
Phys. Rev. Lett. {\bf 83}, 4237 (1999);
L. Pitaevskii and S. Stringari,
Phys. Rev. Lett. {\bf 87}, 180402 (2001).

\bibitem{anglin} J.R. Anglin, P. Drummond, and A. Smerzi,
Phys. Rev. A {\bf 64}, 063605 (2001).

\bibitem{mahmud} K.W. Mahmud, H. Perry, and W.P. Reinhardt,
J. Phys. B: At. Mol. Opt. Phys. {\bf 36}, L265 (2003);
K.W. Mahmud, H. Perry, and W.P. Reinhardt,
Phys. Rev. A {\bf 71}, 023615 (2005).

\bibitem{anna} G. Ferrini, A. Minguzzi, F. W. Hekking,
Phys. Rev. A {\bf 78}, 023606(R) (2008).

\bibitem{cirac} J.I. Cirac, M. Lewenstein, K. Molmer, and P. Zoller,
Phys. Rev. A {\bf 57}, 1208 (1998).

\bibitem{dalvit} D.A.R. Dalvit, J. Dziarmaga, and W.H. Zurek,
Phys. Rev. A {\bf 62}, 013607 (2000).

\bibitem{huang} Y.P. Huang and M.G. Moore,
Phys. Rev. A {\bf 73}, 023606 (2006).

\bibitem{carr} L.D. Carr, D.R. Dounas-Frazer, and
M.A. Garcia-March, EPL {\bf 90}, 10005 (2010).

\bibitem{brand} D.W. Hallwood, T. Ernst, and J. Brand,
Phys. Rev. A  {\bf 82}, 063623 (2010)

\bibitem{main} G. Mazzarella, L. Salasnich, A. Parola, F. Toigo, Physical Review A {\bf 83}, 053607 (2011).

\bibitem{twomode:milburn} C. J. Milburn, J. Corney, E. M. Wright, D. F. Walls, Phys. Rev. A.  {\bf 55}, 4318 (1997).


\bibitem{sb} G. Mazzarella and L. Salasnich, Phys. Rev. A  {\bf 82}, 033611 (2010).

\bibitem{io-e-boris} L. Salasnich, B.A. Malomed, and F. Toigo, Phys. Rev. A {\bf 81}, 045603 (2010).


\bibitem{heidelberg} W. Neuhauser, M. Hohenstatt, P.Tosheck, and H. Dehmelt, Phys. Rev. A {\bf 22}, 1137 (1980).

\bibitem{wineland1} D. J. Wineland, and W. M. Itano, Phys. Lett. {\bf 82A}, 75 (1981).

\bibitem{bergquist} J. C. Bergquist, R. G. Hulet, W. M. Itano, and D. J. Wineland, Phys. Rev. Lett. {\bf 57}, 1699 (1986).

\bibitem{cheinet} P. Cheinet, S. Trotzky, M. Feld, U. Schnorrberger, M. Moreno-Cardoner, S. F\"olling, and I. Bloch, Phys. Rev. Lett. {\bf 101},
090404 (2008).

\bibitem{diedrich} F. Diedrich, J. C. Bergquist, W. M. Itano, D. J. Wineland, Phys. Rev. Lett. {\bf 62}, 403 (1989).

\bibitem{monroe} C. Monroe, D. M. Meekhof, B. E. King, S. R. Jefferts, W. M. Itano, and D. J. Wineland, Phys. Rev. Lett. {\bf 75}, 4011 (1995).

\bibitem{haroche-raimond}  S. Haroche and J. M. Raimond, {\it Exploring the Quantum} (Oxford University Press, Oxford, UK) (2006).

\bibitem{nobel} D. J. Wineland, Rev. Mod. Phys. {\bf 85}, 1103 (2013).

\bibitem{braunstein} W.K. Wootters, Phys. Rev. D. {\bf 23}, 357 (1981); C. W.
Helstrom, {\it Quantum Detection and Estimation Theory} (Academic Press,
New York, 1976), Chap. VIII; A. S. Holevo, {\it Probalistic and Statistical
Aspect of Quantum Theory} (North-Holland, Amsterdam, 1982); S.L. Braunstein and C. M. Caves,
Phys. Rev. Lett. {\bf 72}, 3439 (1994).

\bibitem{pezze} L. Pezz\`{e} and A. Smerzi, Phys. Rev. Lett. {\bf 102}, 100401 (2009).


\bibitem{bwae} C.H. Bennett, H.J. Bernstein, S. Popescu, B. Schumacher,
Phys. Rev. A {\bf 53}, 2046 (1996); S. Hill and W. Wootters,
Phys. Rev. Lett. {\bf 78}, 5022 (1997); L. Amico, R. Fazio, A. Osterloh, V. Vedral, Rev. Mod. Phys.
{\bf 80}, 517 (2008); J. Eisert, M. Cramer, M. B. Plenio, Rev. Mod. Phys.
{\bf 82}, 277 (2010).


\bibitem{cohen} C. Cohen-Tannoudji, B. Diu, F. Laloe, {\it Quantum Mechanics}, Vol. 2, (J. Wiley, New York, 1977).

\bibitem{arecchi} F.T. Arecchi, E. Courtens, R. Gilmore, H. Thomas, Phys. Rev. A {\bf 6}, 2211 (1972).

\bibitem{jaksch} D. Jaksch, H. J. Briegel, J. I. Cirac, C. W. Gardiner, and P. Zoller, Phys. Rev. Lett. {\bf 82}, 1975 (1999).

\bibitem{greiner} M. Greiner, O. Mandel, T. Esslinger, T. W. hansch, and I. Bloch, Nature {\bf 415}, 39 (2002).

\bibitem{oguri} A. Oguri, Phys. Rev. B {\bf 63}, 115305 (2011).






\end{thebibliography}
\end{document}